\documentclass{aa}
\usepackage{txfonts}
\usepackage{psfig}

\newcommand{\ark}{\hbox{Ark~564} }
\newcommand{\arkp}{\hbox{Ark~564.} }
\newcommand{\etal}{et al. }

\newcommand{\asca}{{\it ASCA }}
\newcommand{\xte}{{\it RXTE }}
\newcommand{\xmm}{{\it XMM-Newton }}
\def\simlt{\lower.5ex\hbox{\ltsima}}            
\def\simgt{\lower.5ex\hbox{\gtsima}}            

\def\flux{erg\,cm$^{-2}$\,s$^{-1}$}
\def\la{~\raise.5ex\hbox{$<$}\kern-.8em\lower 1mm\hbox{$\sim$}~}
\def\ma{~\raise.5ex\hbox{$>$}\kern-.8em\lower 1mm\hbox{$\sim$}~}

\begin{document}
\title{Spectral variability analysis of an XMM-Newton observation of Ark~564} 

\author{W. Brinkmann\inst{1} \and I.E. Papadakis\inst{2} \and 
 C. Raeth\inst{1}  } 
\offprints{W. Brinkmann;  e-mail: wpb@mpe.mpg.de}
\institute{Max--Planck--Institut f\"ur extraterrestrische Physik,   
   Giessenbachstrasse, D-85740 Garching, Germany 
\and Physics Department, University of Crete, P.O. Box 2208,
   710 03 Heraklion, Greece  }
 
\date{Received ?/ Accepted ?} 
\abstract
{We present a spectral variability analysis of the X-ray emission of 
 the Narrow Line Seyfert 1
galaxy Ark\,564 using the data from a $\sim$ 100\,ks XMM-Newton observation.}
{Taking advantage of the high sensitivity of this long observation and the
simple spectral shape of Ark 564, we determine accurately the spectral
variability patterns in the source.}
{ We use standard cross-correlation methods to investigate the correlations   
between the soft and hard energy band light curves. We also generated 200   
energy spectra from data stretches of 500\,s duration each and fitted each    
one of them with a power law plus a bremsstrahlung component (for the soft  
excess) and we investigated the correlations between the various best        
fit model parameter values.}
{ The ``power law plus bremsstrahlung'' model
describes the spectrum well at all times. The iron line and the
absorption features, which are found in the time-averaged spectrum
of the source are too weak to effect the results of the time resolved 
spectral fits.  We find that the power law and the soft excess flux
are variable, on all measured time scales. The power law slope is also
variable, and leads the flux variations of both the power law and the 
bremsstrahlung components.}
{Our results  can be explained in the framework of time-dependent
Comptonization models. They are consistent with a picture where instabilities
propagate through an extended X-ray source, affecting first the soft and
then the hard photons producing regions. The soft excess could correspond
to ionized disc reflection emission, in which case it responds fast to the
primary continuum variations.
The time scales are such that light travel times might additionally influence
 the observed variability structure.}

\maketitle
   
\section{Introduction}
\smallskip

Ark~564 is the X-ray brightest Narrow-line Seyfert~1 (NLS1) galaxy with  a
2$-$10~keV flux of $\sim 2\times10^{-11}$ \flux (Turner \etal 2001). From a
recent deep \xmm observation Papadakis et al. (2006, Paper\,I)  found a
hard X-ray power law slope of $\Gamma = 2.43\pm0.03$, in agreement with
previous ASCA (Turner \etal 2001) and BeppoSAX (Comastri \etal 2001) results.
The strong Fe\,K$\alpha$-line claimed by Turner \etal (2001) was not seen.
Instead, a weak ($\sim 90$ eV) emission line, possibly from ionized iron, was
detected. The presence of a soft excess was verified but neither the  existence
of a strong edge-like absorption feature at 0.712 keV (Vignali \etal 2004) nor
the reported emission like feature at $\sim$ 1 keV (Comastri \etal 2001) could
be confirmed. Instead, the soft excess is rather smooth and featureless, with
the most notable feature being a broad,  shallow flux deficit in the energy 
range $0.65-0.85$ keV, similar to the deep troughs that have been detected in
the soft X-ray spectra of several Seyfert galaxies and are inferred to be
``Unresolved Transition Array" (UTA) of iron $n=2-3$ absorption lines. The
smooth soft excess component could equally well be fitted  by a
relativistically blurred photo-ionized disc reflection model or a two black body
components (kT$\sim 0.15$ and $0.07$ keV) model. 

Ark~564 shows large amplitude flux variations on short time scales (Leighly
1999). The source  was observed for a period of $\sim 35$ days in June/July
2000 by \asca as part of a multi-wavelength AGN Watch monitoring campaign
(Turner \etal 2001).Using these data,  Papadakis  \etal (2002) reported a ``-1
to -2" break in the power spectrum at high frequencies ($\sim 2\times10^{-3}$
Hz). On the other hand, Pounds \etal (2001) detected a "zero to -1" low
frequency break in the PSD at $\sim 1/13$ days$^{-1}$, using long term  \xte
monitoring observations. When combined together, these two results support the
idea of a small  black hole mass and high accretion rate in \ark. Finally,
using nonlinear techniques Gliozzi \etal (2002)  were able to demonstrate that
the source behaves differently in the high and low flux states. 

Using the same long \asca monitoring data, Edelson \etal (2002) and Turner
\etal (2001) studied the spectral variability behavior of the source. Edelson
\etal (2002) found that the short-timescale (hours-days) variability patterns
were very similar  across energy bands, and the fractional variability
amplitude was almost independent  of energy. Furthermore, no evidence
of lags between any of the energy bands studied could be found. Turner \etal
(2001) detected significant variations on the hard band spectral slope, by a
factor of $\Delta \Gamma = 0.27$, down to a timescale of approximately a day.
Furthermore, the soft excess flux and shape were also  variable down to
timescales of approximately a day. The power law and soft excess fluxes were
well correlated, with no delays, on time scales longer than a day. 

The integrated luminosity between 10$^{-5}$ and 10 keV of \ark\ is $\sim
10^{45}$ ergs s$^{-1}$ (Romano \etal 2004). Its black hole mass estimate of  
$\sim 2.6\times 10^6 M_\odot$ (Botte \etal 2004) suggests that the system is
accreting near its Eddington limit. Consequently,  the time scales in the
innermost region of the accretion disc around such a black hole are expected to
be  very short. For example the light travel time scale is of the order of a
few $\times$ 100 s in the innermost $\sim$10 Schwarzschild radii for such a
small black hole mass. Therefore, in order to probe the physical processes that
operate in the innermost accretion flow we need to study the spectral
variations of the source on time scales much smaller than a day. The high time
resolution and good signal to noise data that \xmm provides are well suited
for this kind of work.  

In this paper we present the results from the time resolved spectral analysis 
of the data from the January 2005, 100 ks \xmm observation of \arkp  Our main
aim is to study the variability behavior of the  continuum components in the
X-ray spectrum of \ark, i.e. of the soft excess and the    hard band power law
component. Ark\,564 is an ideal target for the investigation of the continuum
spectral variability on short time scales. The main reason is that, as we have
shown in Paper I (where we presented the results from the analysis of the
time-averaged spectrum of the source), its X-ray spectrum is almost ``clean" of
any significant absorption and/or emission features. Hence, it is possible to
determine the X-ray continuum components with the use of simple
phenomenological models, without significant uncertainties associated with the
presence of features due to (warm and/or cold) absorbing/emitting material in
the source.                                                

A detailed timing analysis, focusing on the study of the phase lags and the
coherence as a function of Fourier frequency, has been presented  by  Arevalo
\etal (2006), while McHardy \etal (2006) will present the result from a
detailed timing analysis based on power spectral analysis methods. 

The present work is organized as follows:  After a brief description of the
observation we present in Sect.\,3 the  timing analysis of the light curves,
both with conventional techniques as well as with a sliding window approach. 
In section 4 we describe the  time resolved spectral fits and the correlations
of the obtained  spectral parameters as well as the correlations of the fluxes
of the different components in different energy bands.  We close in section 5
with a discussion of the obtained results.

\section{Observation and Data Analysis}

Ark~564 was observed with \xmm from 2005 January 5, 19:47 to 2005 January 6,
23:16 for about 10$^5$\,s. The PN camera and  the two MOS cameras were operated
in Small Window mode with a medium thick filter. For the current analysis we
will use only the PN data.  The data were reprocessed  with the XMMSAS  version
6.5 and for the spectral analysis the most recent versions of the response
matrices were used.

The background count rate was very low (in total less than 0.6\% of the source 
count rate), apart from two small, short flares at the beginning of the
observation. Data from these periods were disregarded from the spectral
analysis.

With an average  count rate of $\sim$ 30 cts~s$^{-1}$ photon pile-up is
negligible for the PN detector, as  was verified using the XMMSAS task
$epatplot$. Source counts were accumulated  from 27$\times$26 RAW pixels 
around the  position of the  source;  the background data were extracted from a
similar, source free region on the chip. We selected single and double  events
for the analysis (PATTERN$\leq 4$ and FLAG=0; for details of the instruments
see Ehle et al. 2005) in the energy range from 300 eV to 12 keV. In total
$\sim$ 2$\times10^6$ photons were accumulated in an integration time of $\sim$
69300\,s.

\section{Light curve timing analysis}

\begin{figure}
\psfig{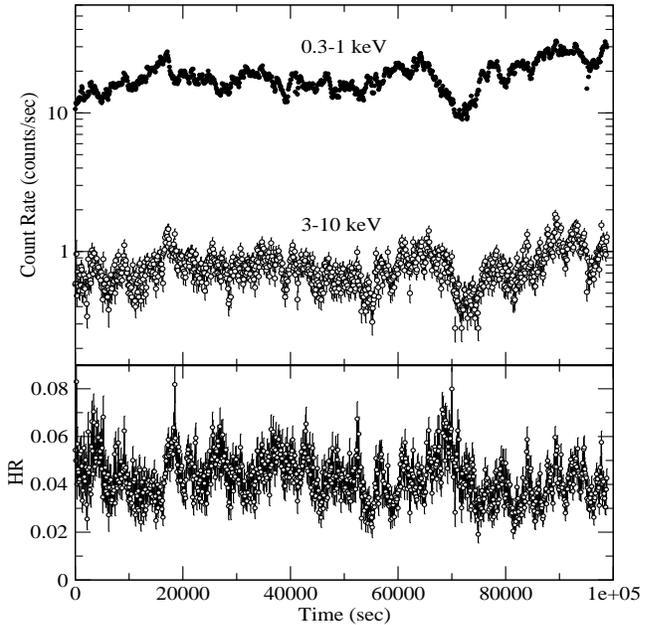}
\caption[]{Background subtracted PN light curves of Ark 564 in the 
0.3$-$1\,keV  (filled dots) and 3.0$-$10\,keV (open circles) energy bands. 
Note that in reality the count rates are 
larger, as the data have  not been corrected for the 71\% life
time of the Small Window mode of the detector. 
Bottom panel: hardness ratios between hard and soft band light curves. 
The time binning is 100\,s.}
\label{fig:lc}
\end{figure}

We start our study by examining light curves in ``soft" and ``hard" bands. The
advantage of this approach is that it is model independent, however it has also
its limitations, as it is difficult to disentangle the interplay between the
various spectral components in each energy band.   Fig.~\ref{fig:lc} shows the 
PN 0.3$-$1\,keV and 3$-$10\,keV background subtracted light curves,  binned in
100\,s intervals. We  consider the light curves in 3$-$10\,keV and 0.3$-$1\,keV 
as representative of the ``hard" and ``soft" energy bands, respectively. The
high energy band range has been chosen to minimize the contribution of the soft
excess emission component which becomes apparent at energies below $\sim 2$ keV
(Paper\,I). The contribution of this component is significant in the soft energy
band light curve. However, as  the hard power law very likely continues into
this band,  the interpretation of the observed variations in this band is
difficult.

The source is highly variable on all sampled time scales in both energy bands. 
The max/min variability amplitude is of the order of $\sim 6$ and 4 in the hard
and soft band light curves, respectively. The fractional variability amplitude
(corrected for the experimental noise) in the  hard energy band  is  $f_{\rm
rms, hard}=27.1 \pm$ 0.4\%, as opposed to  $f_{\rm rms, soft}
=24.7\pm 0.1$\% in the soft band light curve. Errors account only for the
measurement error of the light curve points, and have been estimated according
to the prescription of Vaughan \etal (2003). This difference in the variability
amplitudes suggest the presence of spectral variations which can be seen in the
lower panel of  Fig.~\ref{fig:lc} where we plot the hardness ratios (HR), i.e.
the ratio of the hard to the soft band light curves. A $\chi^{2}$ test shows
that the HR light curve is indeed significantly variable ($\chi^{2}=3464/988$
degrees of freedom - dof). The fractional variability amplitude of the HR light
curve is  $f_{\rm rms,HR} =17.9\pm 0.4$\%, suggesting that the amplitude of
the  spectral variations is lower than that of the flux variations in both the
soft and hard band light curves. 

\subsection{Cross-Correlation Analysis}

In order to investigate the cross-links between the hard and soft band light
curves we estimated their cross-correlation function (CCF), using light curves
with a 100-s bin size.  We calculated the sample Cross-Correlation Function,
$CCF(k)$, as follows:

\[
CCF(k)=\frac{\sum_{t}(x_{soft}(t) -\bar{x}_{soft})(x_{hard}(t+k)-\bar{x}_
{hard})}{N(k)(\sigma^{2}_{soft}\sigma^{2}_{hard})^{1/2}}, \]

\[ k=0,\pm \Delta t,\ldots,\pm (N-1)\Delta t. \]

The summation goes from $t=\Delta t$ to $(N-k)\Delta t$ for $k\geq 0$ ($\Delta
t = 100$ s, and $N$ is the total number of points in the light curve)  and the
sum is divided by the number of pairs included, i.e. $N(k)$.  For negative lags
$k < 0$ the summation has to be done over x$_{\mathrm soft}(t +|k|)$ and
x$_{\mathrm hard}(t)$. The variances in the above equation are the source
variances, i.e., after correction for the experimental variance. Significant
correlation at positive lags means that the soft band variations are leading
those in the hard band.

\begin{figure}
\psfig{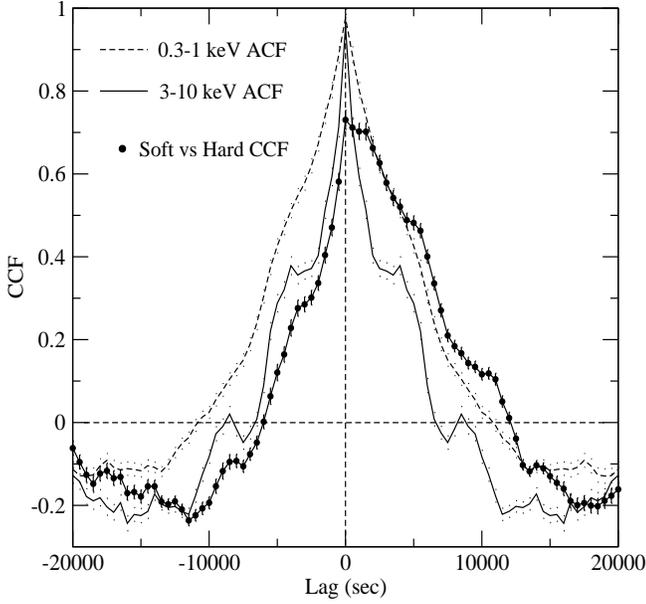}
\caption[]{\small Autocorrelation functions of the soft and hard 
 band data  and the cross-correlation function between the soft and the 
 hard light curve. The time binning is 100\,s. Positive lags mean
 that the hard variations are lagging behind the soft.} 
\label{fig:ccftot}
\end{figure}
 
\begin{figure}
\psfig{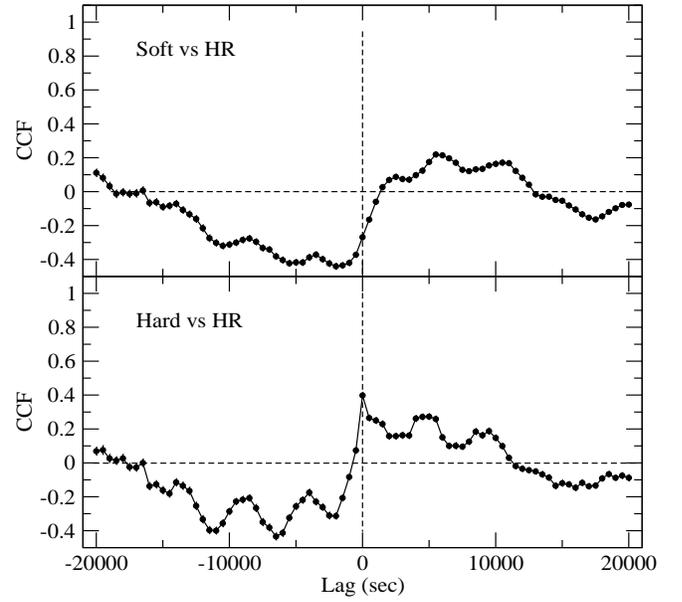}
\caption[]{\small Cross-correlation function between the   HR and the the soft
- hard band light curves (upper and lower panels, respectively).The time
binning is 100\,s. Positive lags mean that the HR spectral variations are
lagging behind the flux variations.}
\label{fig:ccfhr}
\end{figure}
 
First, we computed the auto-correlation function (ACF) of the hard and soft
band light curves and the CCF (up to lags $\pm 2\times10^4$ s) using the  whole
length (i.e. the total) light curves.  In Fig.\ref{fig:ccftot}
we show the soft and hard band ACFs (continuous and dashed lines,
respectively)  and their cross-correlation function (filled circles). 

The  maximum CCF amplitude reaches a value of $\sim$ 0.7, at  zero lag. This
result implies that the  soft and the hard band light curves are well
correlated. This is not surprising, given the fact that the two light curves
look very similar (see Fig.~\ref{fig:lc}), with most of the variability
patterns, on all time scales, appearing in both of them. Furthermore, the CCF
appears to be highly asymmetric with higher values appearing at positive lags.
The asymmetry of the hard vs soft CCF becomes immediately apparent if we compare
it with the respective ACFs. This result  implies complex delays between the
hard and the soft band light curves. It is in agreement with the results of
Arevalo  \etal (2006) who also find that the variability components in the
0.7$-$2 keV are leading those in in the 2$-$10 keV band. However, the time delay
between the variations in the two bands is not constant, but rather decreases
with increasing period of the Fourier components. In this case, broad,
asymmetric towards positive lags CCFs are expected,  like the one shown in Fig.
\ref{fig:ccftot}. 

The cross-correlation between the soft and hard band light curves and the
hardness ratios (Fig.\ref{fig:ccfhr}) is rather weak at all lags. The
``strongest" signal appears at negative lags which implies that the flux
variations may be weakly anti-correlated with the HR variations with a delay of 
$\sim 2-$5\,ks. If real, this  anti-correlation between the soft and hard band
fluxes and the hardness ratio is hard to explain given the possible contribution
of both the strong soft excess and the power law component in the soft band
light curve of this object.  

\subsection{Sliding window CCFs}

The CCFs shown in  Figs. \ref{fig:ccftot} and \ref{fig:ccfhr} represent the
average correlation structure between the two light curves, originating from
the integration over the total length of the observation. Recently, Brinkmann
\etal (2005) employed the ``sliding window"  technique in the study of the CCFs
between various energy bands in the case of the BL Lac object Mrk~421. As is
typical of objects in the same class, Mkn~421 shows well defined flare-like
events in its light curves, with different variability characteristics. As a
result, the use of this method revealed that the cross-correlation structure
also evolves during each individual events.

In order to investigate in more detail the cross-correlation between the soft
and hard band variations, over shorter time periods we employed the same 
``sliding window" technique and calculated the CCFs for shorter data intervals,
which start at different times of the observed light curve and range over a
restricted time interval. Thus, the above defined CCF(k) is replaced by a CCF(k;
$\mathcal{T,L}$), i.e., the cross-correlation coefficient at a lag `k' is
calculated for a data stream with length $\mathcal{L}$ which starts at the time
$\mathcal{T}$ in the light curve. Details of the method and the choice of an 
optimal window length $\mathcal{L}$ which reveals best the ``time evolution" of
the CCF structure are given in  Brinkmann \etal (2005). 

\begin{figure}
\psfig{figure=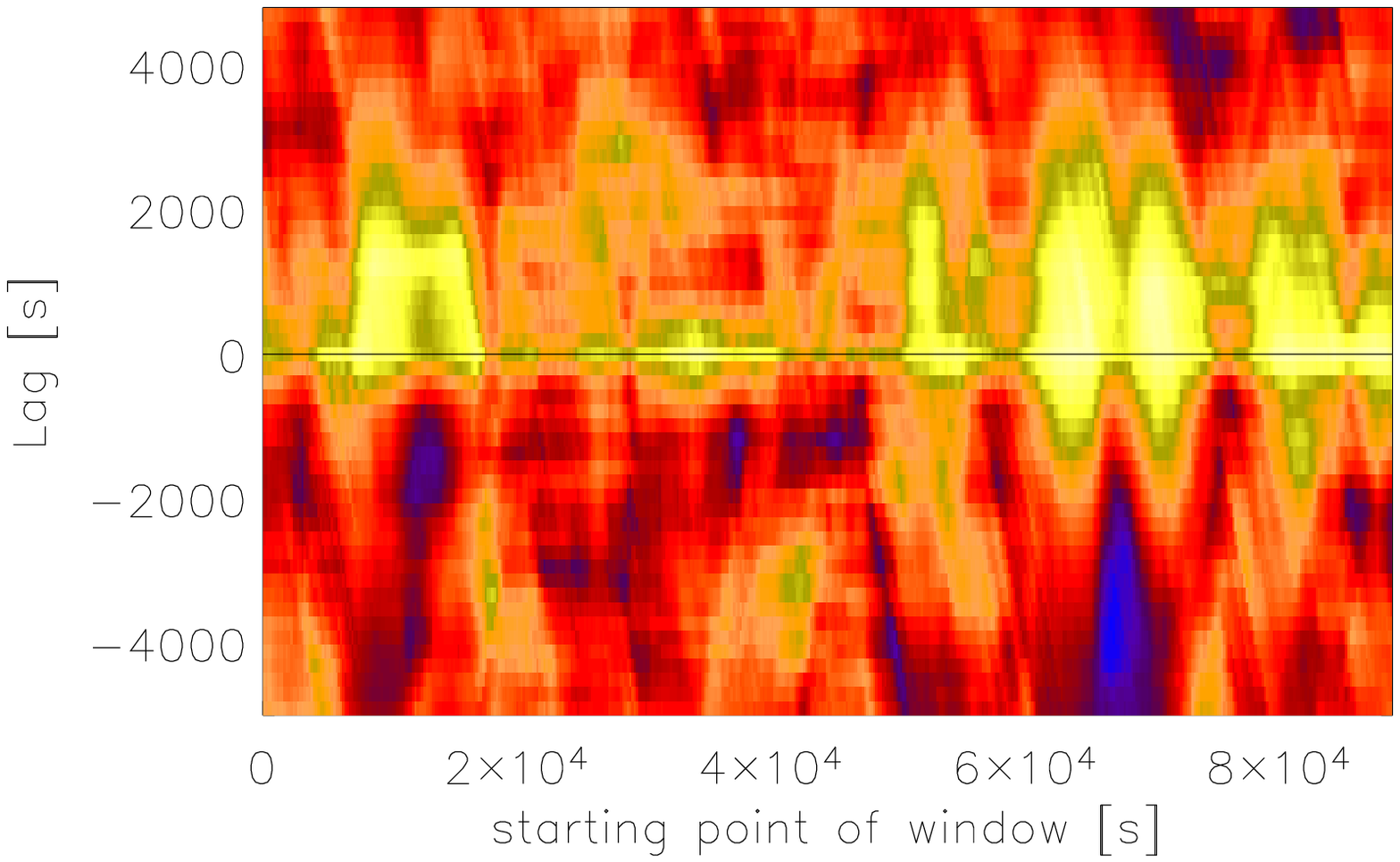,height=4.3truecm,width=8.5truecm,angle=0,%
 bbllx=2pt,bblly=1pt,bburx=490pt,bbury=290pt,clip=}
\psfig{figure=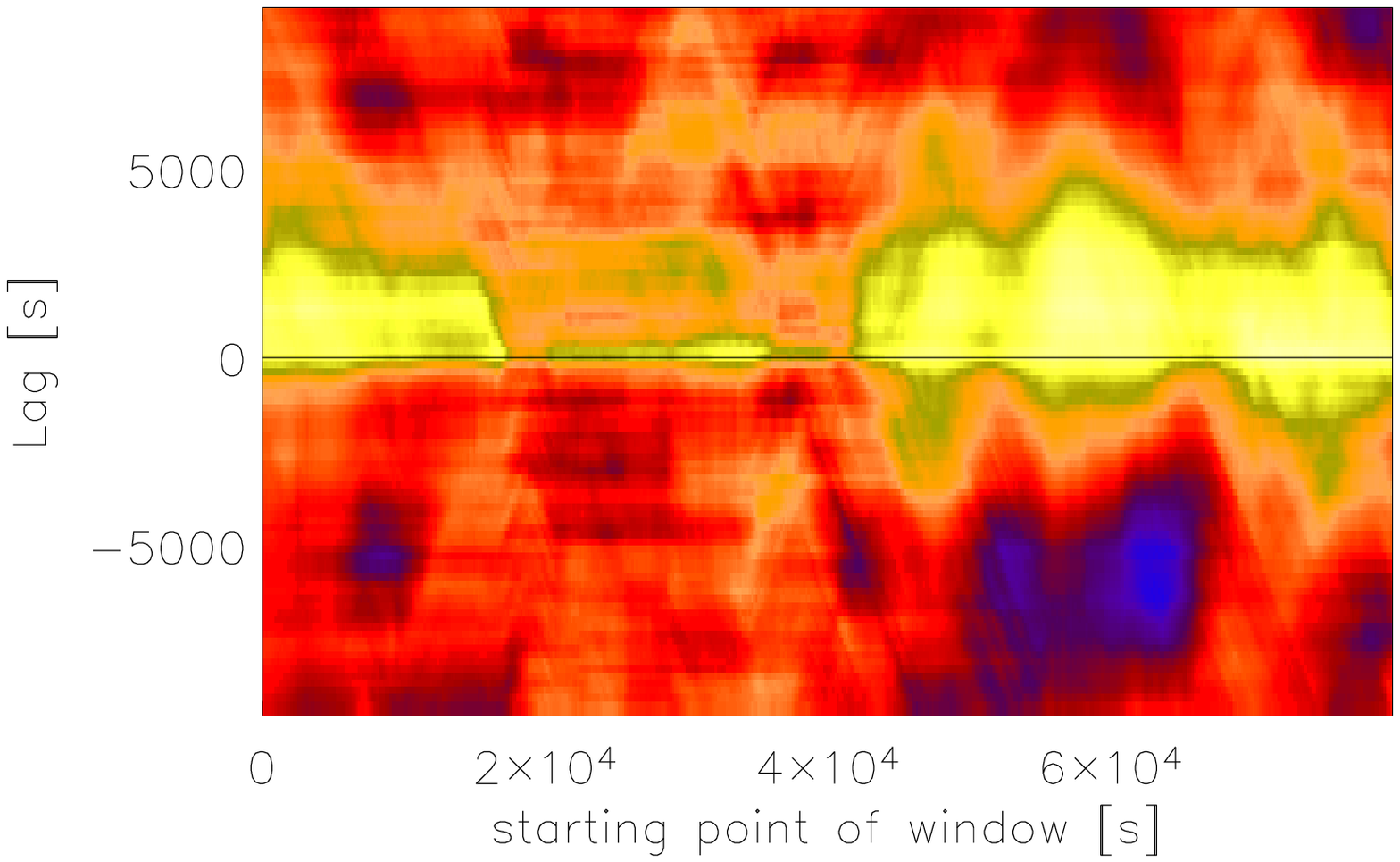,height=4.3truecm,width=8.5truecm,angle=0,%
 bbllx=2pt,bblly=1pt,bburx=490pt,bbury=290pt,clip=}
\psfig{figure=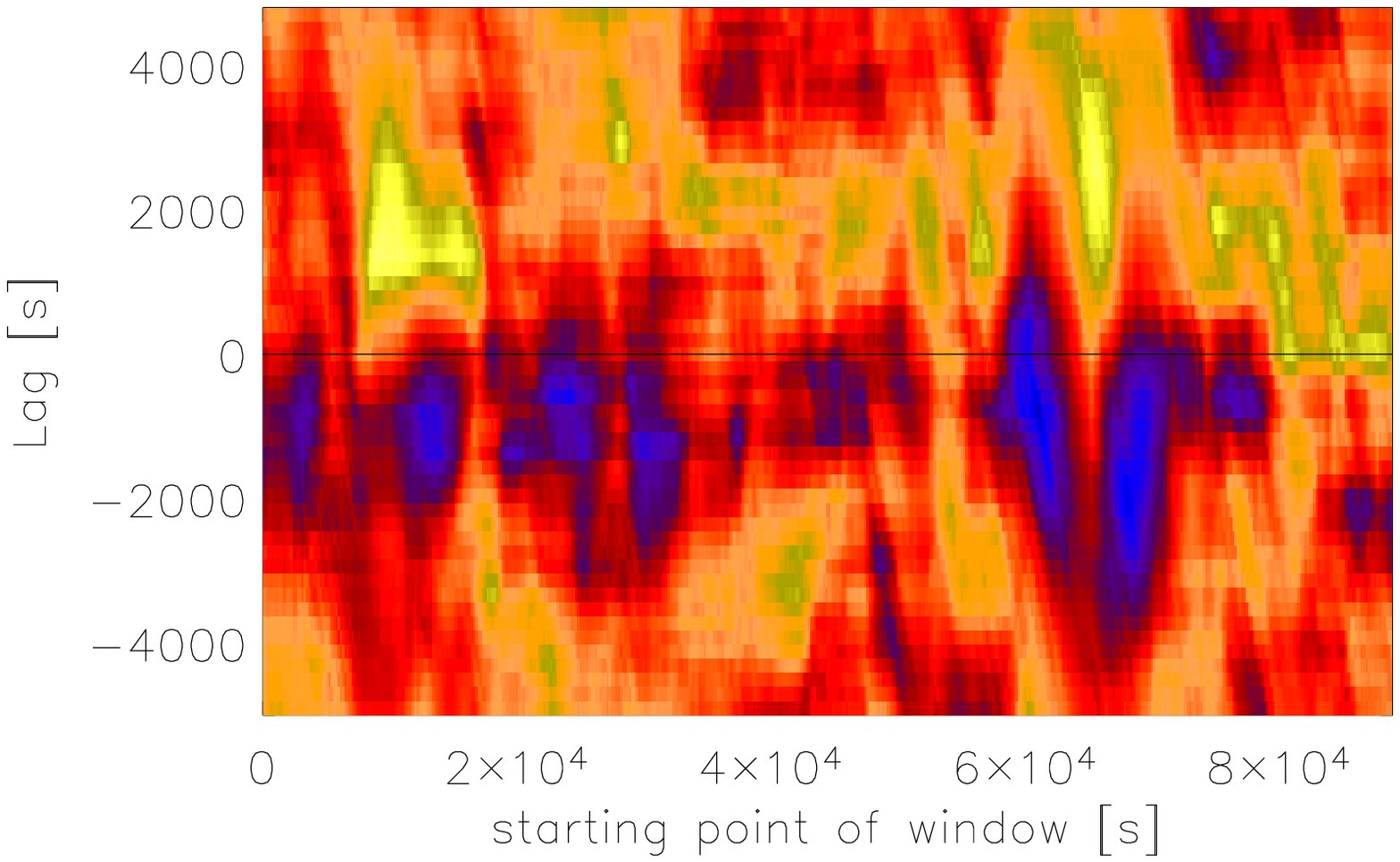,height=4.3truecm,width=8.5truecm,angle=0,%
  bbllx=2pt,bblly=1pt,bburx=490pt,bbury=290pt,clip=}
\psfig{figure=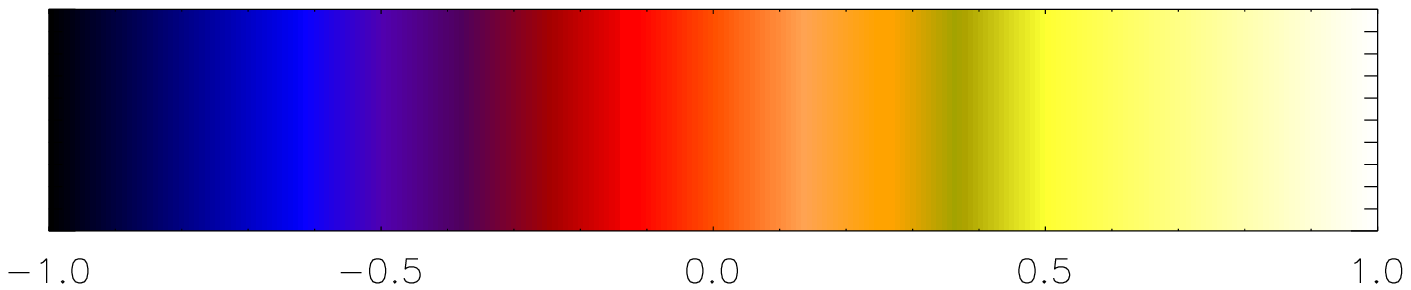,height=1.8truecm,width=8.5truecm,angle=0,%
 bbllx=11pt,bblly=1pt,bburx=485pt,bbury=85pt,clip=}
 \caption[]{\small Sliding window CCFs between soft and hard band fluxes  (upper
panel with window length $\mathcal{L}$ = 10\,ks, middle panel  with  window
length $\mathcal{L}$ = 19\,ks). In the lower panel we plot the sliding window
CCFs ($\mathcal{L}$ = 10\,ks) between the soft  flux and the hardness ratios. 
The amplitude of the CCF is color coded,  with lags plotted in the vertical
direction. The bottom panel shows the color bar used for the numerical values
of the  CCFs.}   
\label{fig:slide1}
\end{figure}

In Fig. \ref{fig:slide1} we show a two dimensional representation of the
sliding window CCFs, based on light curves with 10\,s binning. The vertical
scale represents the lags (similar to the x-axis of  Figs.~\ref{fig:ccftot} and
\ref{fig:ccfhr}), the color coding indicates the amplitude of the
cross-correlation coefficient, with the color code given in the lowest panel of
the figure. Time is along the x-axis.  The two upper panels show the
cross-correlation function between the soft and hard band light curves for two
window lengths, $\mathcal{L}$ = 10 ks and $\mathcal{L}$ = 19 ks. 

Clearly visible is the temporal sub-structure in the CCFs, which gets averaged
out for the longer window lengths and approaches, finally, the ``global''
averages seen in Fig.\ref{fig:ccftot}. During the period of ``low-amplitude''
activity (i.e. between $\sim 20$ and $\sim 60$ ks after the start of the
observation) the CCFs are of lower amplitude and peak around zero lag. During
the first and last part of the  observation though, the CCF has a larger
amplitude and  peaks at positive lags, between $\sim 0-2$ ks.  The main result
of this analysis is that there exist  strong correlations only at zero or
positive lags. This is in contrast to, for example, Mrk~421 where lags with 
changing signs have been observed (e.g. Brinkmann \etal 2005).

The third panel of  Fig. \ref{fig:slide1} shows the sliding window CCFs between
the soft band flux and the hardness ratios. The length of the data sets is again
$\mathcal{L}$ = 10~ks. Overall, the  correlation between these quantities is
rather low, at all times. During the ``low-amplitude" activity period (i.e.
during the middle of the observation)  we cannot detect any correlation between
the two light curves, most probably due to the low amplitude variations. 
Towards the end and at the beginning of the observation, we detect the signals
at positive and negative lags of  $\sim 2$ and $\sim -2$ ks, respectively. In
both cases though, the positive and negative lag ``signals" are of low amplitude
(i.e. CCF$_{\rm max}\la 0.5$). 

In order to proceed and to investigate further the correlation between the flux
and spectral variations in \ark\, we need to disentangle the contribution of the
hard and soft spectral components  in the soft energy band. This can be achieved
by model fitting to the energy spectra acquired over short periods of time. In
the section below we present the results from this temporally resolved spectral
analysis. 

Before proceeding, it is worth mentioning that one of the main results from the
``sliding window" CCF analysis is that the ``global" cross-correlation
properties between the light curves in the two energy bands are mainly defined
by the ``signals" we detect at the beginning and towards the end of the
observation  (i.e. during the first $\sim 20$ ks and in the period  between 60
and 80 ks after the start of the observation). In the first period we observe a
steady flux increase (by a factor of $\sim 2-3$) and then a much shorter flux
decline (over a period of less than $\sim 2-3$ ks), reminiscent of a flare-like
event. In the second period, we observe a long ($\sim 10$ ks), well defined flux
decay (by a factor again of $\sim 2-3$) and a subsequent flux rise over a
similar time scale. These are the most prominent, long, well defined ``features"
in the observed light curves  (see Fig.~\ref{fig:lc}). In the periods between
these two main ``events", the flux variations are of lower amplitude and are
less well defined. Hence  it is harder to detect strong cross-correlation
signals. In the analysis presented in the following sections, we pay particular
attention to these two periods which show the strongest cross-correlation
signals between the soft and hard band light curves.

\section{Time resolved spectral fitting analysis}
\smallskip
  
\begin{figure}
\psfig{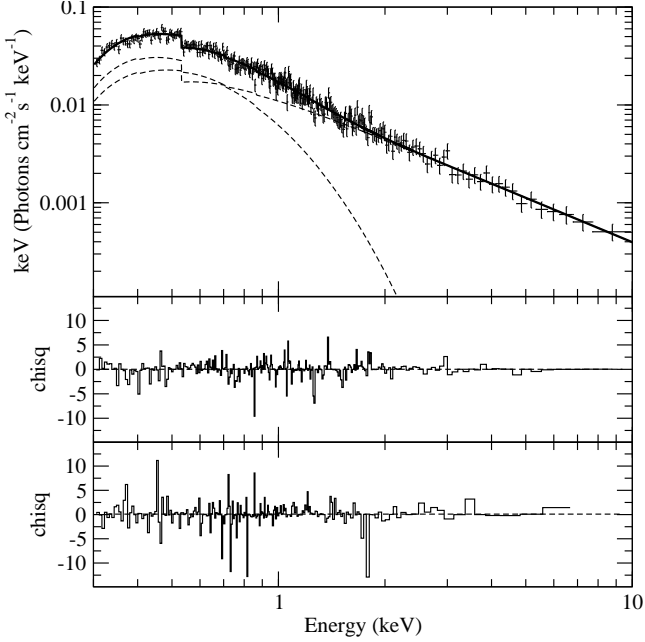}
\caption[]{\small Plot of a good power law plus bremsstrahlung
 fit (upper panel) and of the respective residuals (middle panel) 
   to the PN data in the $0.3-10$ keV energy band. The lower panel
shows the residuals of the worst fit obtained.}
\label{fig:pn-test}
\end{figure}

As the count rate of the \ark \xmm observation is sufficiently large, we
split up the total exposure into  200 individual data stretches of 500\,s
duration each, and generated the energy spectrum in each of them.  Each of the
thus obtained spectra has $\ma$ 10000 photons in the $0.3-10$\, keV band,
sufficient for the accurate determination of the parameters of the relatively
simple spectral models that we use. All model fits were done using XSPEC v.
11.3, and in all cases we fixed the interstellar absorption at the   galactic
value of N$_{\rm H, Gal}$ =   $6.4\times10^{20}$ cm$^{-2}$ (Dickey \& Lockman
1990). In all cases we use the 1$\sigma$ errors for one interesting parameter.
 
A simple power law (PL) model fits the $3-10$ keV band of all these  spectra
very well. However, it cannot provide an acceptable fit to the full
$0.3-10$ keV band.  A soft excess emission component is always present at
energies below $\sim 1.5$ keV. We investigated whether a combination of just
two spectral components could fit the entire $0.3-10$ keV band. We used a
combination of a PL plus a black body or a PL plus a bremsstrahlung  
(PL+Brems) model and found that the second combination fits most of the     
500\,s spectra significantly better.  It results nearly always in an        
excellent fit: the mean reduced $\chi^2$ of all 200 fitted spectra is       
$<\chi^2_{\rm red}>$ = 1.0024, with a variance of $\sigma = 0.0068$.

In Fig.\,\ref{fig:pn-test} we show one of the best PL+Brems fits to one of  
our 500\,s data stretches ($\chi_{\rm red}^2 =$ 0.951 / 279 dof) and the    
corresponding $\chi^2$ plot (top and middle panel, respectively). In the   
bottom panel, we plot the $\chi^2$ plot of the worst fit we obtained        
($\chi_{\rm red}^2 =$1.31 / 217 dof). In both cases the resulting quality   
of the fits is determined by the scatter of individual data bins. There     
are no obvious systematic deviations between data and the model, 
suggestive of the presence of any extra spectral components that we should  
consider.
  
In particular, none of the absorption and emission features that were       
detected in the time-averaged spectrum of the source (i.e. the weak iron     
emission line at $\sim 6.7$ keV, the absorption line at $\sim 8$ keV, and   
the broad, but shallow, flux deficit at $\sim 0.7-0.9$ keV; see Paper I)    
showed up in the spectra which are extracted from the shorter data          
intervals, due to the limited photon statistics.
   
In order to investigate this issue better, we fitted again the lowest,  
highest flux, and the worst-fit 500 sec spectra with a PL+Brems model plus  
1) a \footnotesize{DISKLINE} and Gaussian absorption line component (using  
the model parameter values listed in the PN, \footnotesize{PL+DL+ABL}       
model fitting results of Table 1 in Paper I) and 2) the four Gaussian lines    
that we used to model the absorption in the Fe~UTA region.     
The best fitting PL and Brems model results were almost identical to the    
values we obtain when we do not consider the small amplitude  
absorption/emission features in the time-averaged spectrum of the source.
   
We repeated the same procedure for 20 more, randomly chosen 500-sec long    
spectra. In all cases, the addition of model components that correspond to  
the absorption and/or emission features in the time-averaged spectrum of     
the source does not affect at all the continuum (i.e. PL and Brems) model   
fitting results.We conclude that the PL+Brems model can fit well the time
resolved spectra, over the entire 0.3$-$10 keV band. Consequently, we decided
to use the results from the PL+Brems model fitting to the spectra as the basis
for our study of the spectral variability properties of the source.

\begin{figure*}
\hbox to \hsize{ 
\psfig{figure=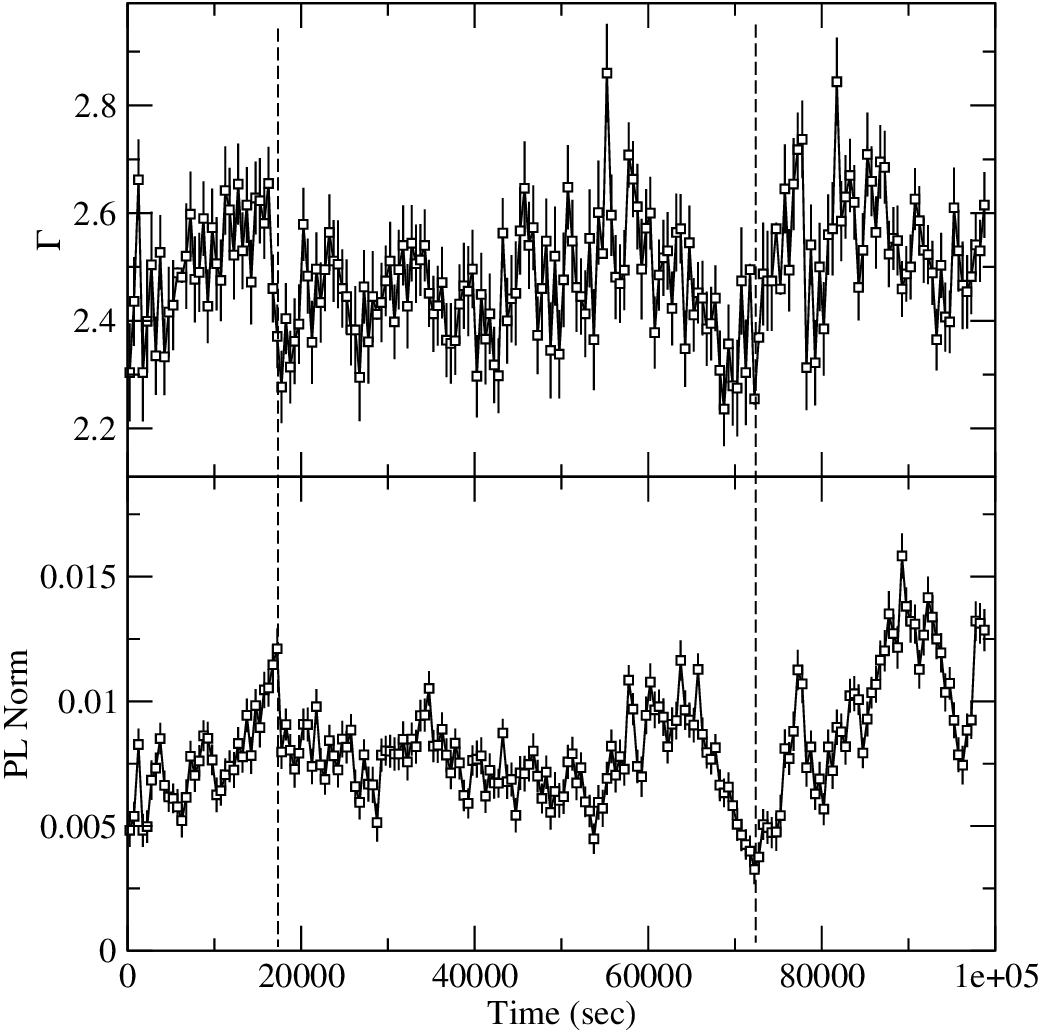,height=9.3truecm,width=8.5truecm,angle=0,%
 bbllx=30pt,bblly=259pt,bburx=527pt,bbury=755pt,clip=}
\psfig{figure=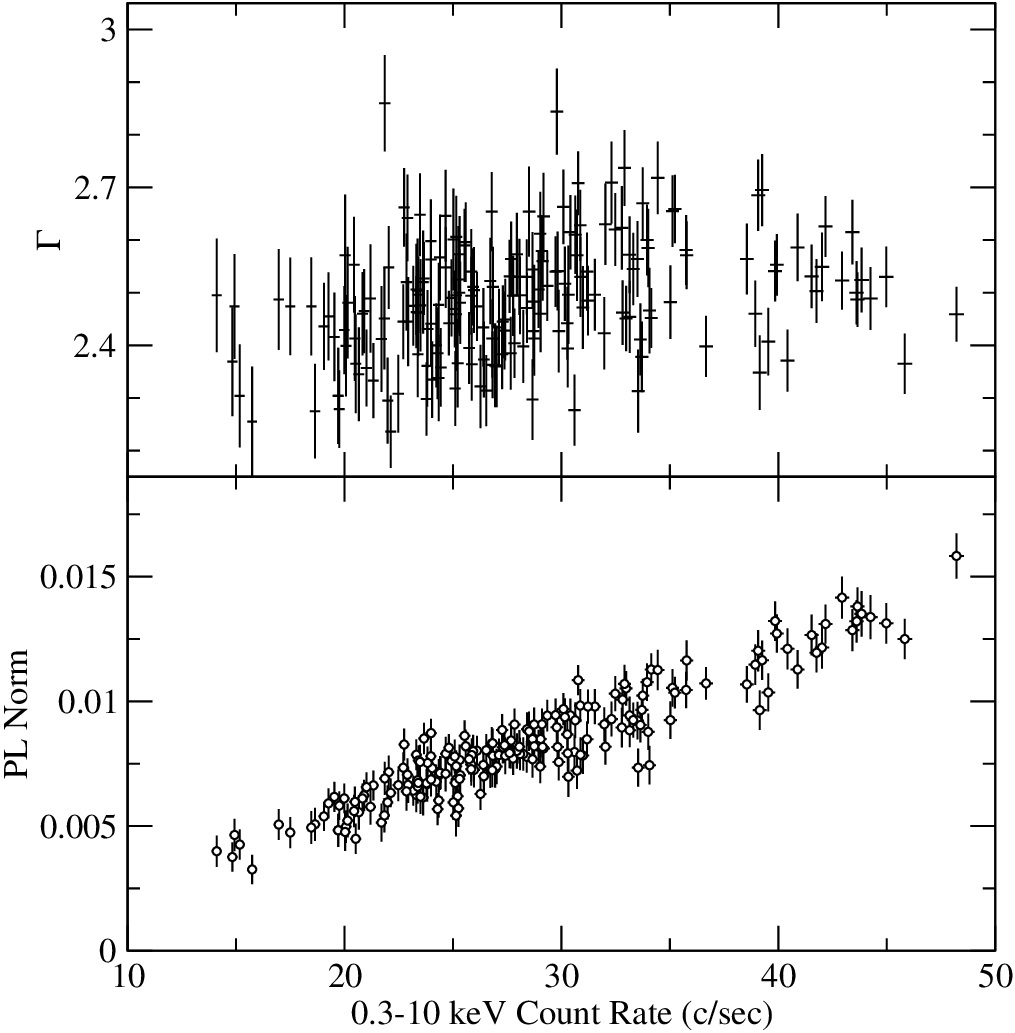,height=9.3truecm,width=8.5truecm,angle=0,%
 bbllx=30pt,bblly=259pt,bburx=517pt,bbury=755pt,clip=}      
 }
\caption[]{\small Left: Best fitting power-law index (upper panel) and
   power-law normalization (lower panel) values plotted as function of the 
   observation time. Right: The same parameters are plotted as a function of
   the 0.3$-$10 keV count rate.}
\label{fig:gamfit}
\end{figure*}
    
\begin{figure}                                                                                                           
\psfig{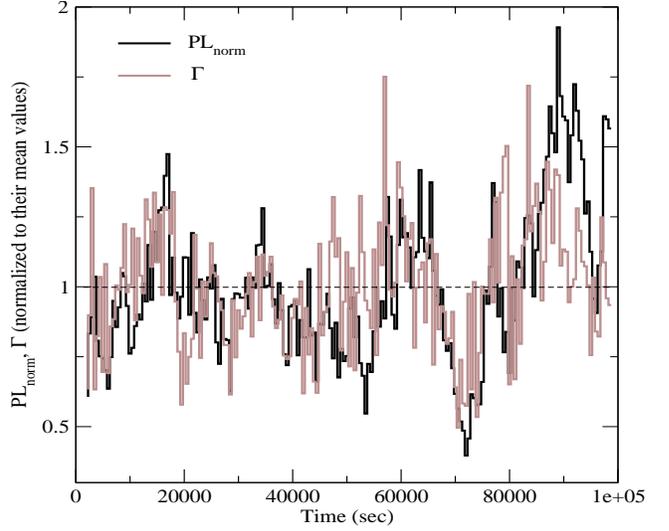}                   
 \caption[]{\small Power law normalization (black) and 
 and amplitude adjusted power law slopes (grey),  both
 normalized to their mean values, shifted in
 time by +\,2\,ks.}
\label{fig:corr}
\end{figure} 

\subsection{Best fitting power-law parameter correlations}

The left panel of Fig.\,\ref{fig:gamfit} displays  the best
fitting values for the PL spectral slope and normalization as function of the
observing time. In the right panel of the same figure, we show these
parameter values as a function of the  0.3$-$10\,keV count rate. 

The PL normalization (PL$_{\rm norm}$) varies substantially, and these
variations are obviously the primary cause of the sources' variability. As the
right hand bottom panel of Fig.\,\ref{fig:gamfit} confirms, there exists a 
very strong, linear correlation between the observed count rate and PL$_{\rm
norm}$. The max-to-min variability amplitude of $\sim$ 4.8 in the observed
light curve can almost entirely be explained by the PL$_{\rm norm}$ 
variations. 

The power law spectral slope is also variable, but at a smaller level. Its
average value is $<\Gamma> = 2.488\pm0.005$,  similar to best PL model fitting
to the hard band of the time-averaged spectrum, reported in Paper\,I. The
individual PL slopes vary between  $\sim 2.3-2.7$, with an average amplitude of
$f_{\rm rms, \Gamma}=3.3\pm 0.2$\%, much smaller than that of the PL$_{\rm
norm}$ ($f_{\rm rms, PLnorm}=26.0\pm 0.6$\%). On the top right hand panel, we
plot $\Gamma$ as a function of total observed count rate.  There might exist a
weak correlation, possibly with a slightly different behavior at highest
fluxes, but the large uncertainty associated with the $\Gamma$ measurements 
does not allow us to draw any firm conclusions. 

\begin{figure*}
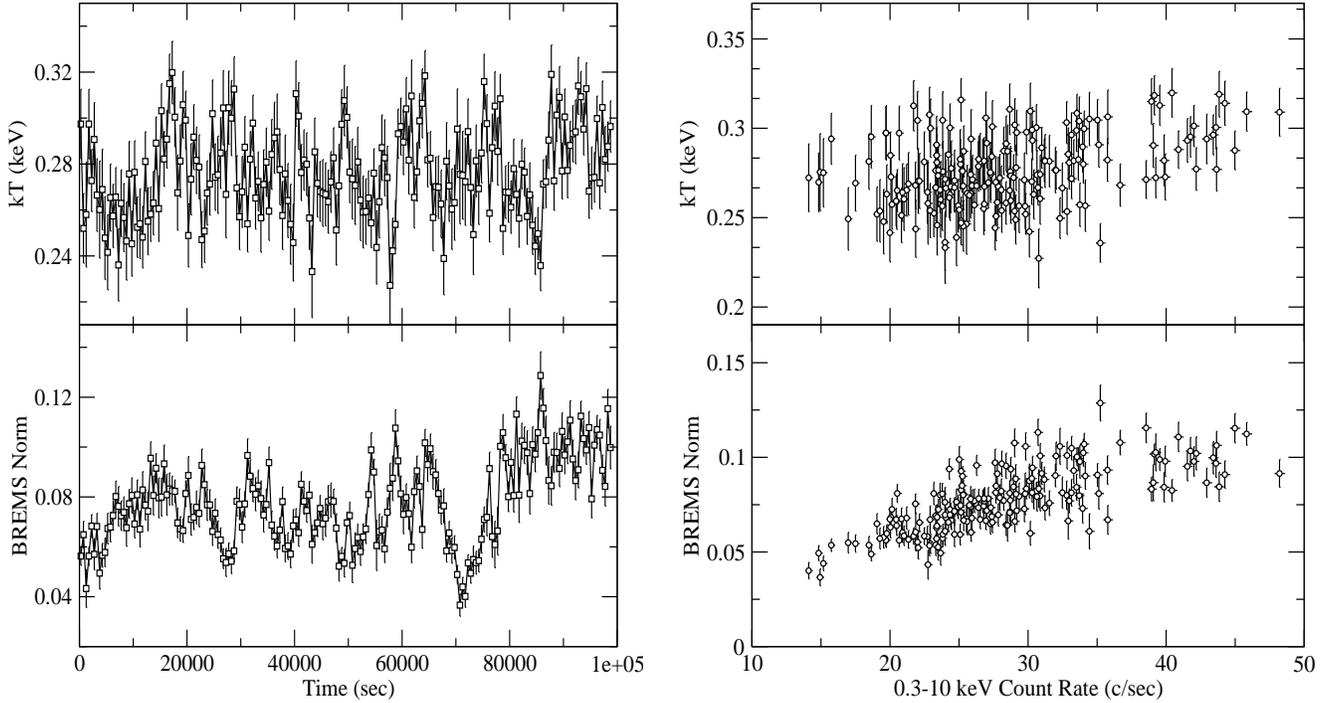

\hbox to \hsize{
\psfig{figure=APH-fig8a.eps,height=9.3truecm,width=8.5truecm,angle=0,%
 bbllx=30pt,bblly=259pt,bburx=527pt,bbury=755pt,clip=}
\psfig{figure=APH-fig8b.eps,height=9.3truecm,width=8.5truecm,angle=0,%
  bbllx=33pt,bblly=259pt,bburx=517pt,bbury=755pt,clip=}
 }
\caption[]{\small Left: Best fitting temperature (upper panel) and the
corresponding normalization values (lower panel) of the bremsstrahlung
component plotted as function of the observation time. Right: The same
parameters plotted as function of the 0.3$-$10 keV count rate.}  
\label{fig:kTfit} 
\end{figure*} 

Despite the difference in the variability amplitudes, $\Gamma$ and PL$_{\rm
norm}$ are well correlated. Interestingly, there also appears to be a time
delay between the variations of these two parameters.  As an illustrative
example  we have drawn two dashed lines in the left hand panels of
Fig.\,\ref{fig:gamfit}. They indicate the high and low PL$_{\rm norm}$ values
$\sim 17$ ks and 70 ks after the beginning of the \xmm observation,
respectively.  The associated $\Gamma$ variations are clearly leading those in
the PL$_{\rm norm}$ light curve, by about  1-2\,ks. 

The good correlation between PL$_{\rm norm}$ and $\Gamma$ is further
demonstrated in Fig.\,\ref{fig:corr} which  shows  $\Gamma$   and  PL$_{\rm
norm}$ plotted as a function of time. Both curves are normalized to their
mean.  The normalized $\Gamma$ values are furthermore scaled by a factor of 5,
and shifted by +\,2\,ks.  Visible is the overall excellent correlation between
the two quantities.  However, there also exist some subtle anti-correlations
and mismatches, for example at  $\sim 20, 45-55,$ and 77 ks after the
observations start. Some of them could be the result of the fact that the delay
between the variations in the two parameters is not exactly equal to 2\,ks.
Towards the end of the observation ($\ma$ 85 ks since its start), the
correlation ceases to be good, with a prominent ``flare-like" event in the
PL$_{\rm norm}$ vs time plot, which is almost absent in the respective $\Gamma$
vs time curve.

\subsection{Best fitting bremsstrahlung parameter correlations}

Figure \ref{fig:kTfit} shows the corresponding results  of the bremsstrahlung
component. On the left hand side panels, we plot  the best fitting kT (upper
panel) and bremsstrahlung normalization values (brems$_{\rm norm}$; lower
panel) as function of time. In the right hand panels, we plot the same
parameter values as a function of the total count rate. 

The bremsstrahlung normalization is variable, with an average variability
amplitude of $f_{\rm rms, brems-norm}=20.0\pm 0.6$\%. This result shows clearly
that the soft excess component {\it is} variable in \arkp In fact, our results
demonstrate that this component is variable on time scales as short as a few
hundred seconds. Its variability amplitude is smaller than that of the PL$_{\rm
norm}$, and of the observed full band light curve. Nevertheless, as with
PL$_{\rm norm}$, brems$_{\rm norm}$ follows rather closely the intensity
variations. The correlation between  brems$_{\rm norm}$ and count rate (as
implied by the right hand bottom panel in  Fig. \ref{fig:kTfit}) is rather good,
although  with substantial scatter, indicative of the strong influence of the PL
component variations in the intensity variations that we observe. There is also
an indication that at the highest count rate level the normalization of the soft
component  may saturate at a constant value. 

Visual inspection of the bottom left hand panels in Figs. \ref{fig:gamfit} and 
\ref{fig:kTfit} reveals that brems$_{\rm norm}$ and PL$_{\rm norm}$ are well
correlated. The cross-correlation analyzes of the two light curves in these
panels does indeed show a maximum of $\sim 0.7$ at a lag of $\sim 0.5-1$\,ks,
indicating that the  PL$_{\rm norm}$ variations may be slightly delayed with
respect to the brems$_{\rm norm}$ variations. Furthermore, the visual
inspection of the top and bottom left hand panel in Figs. \ref{fig:gamfit} and
\ref{fig:kTfit}, respectively, suggests that brems$_{\rm norm}$ is also
correlated with $\Gamma$. A cross-correlation analysis results in a CCF$_{max}$
value of 0.5, at a lag of $\sim 1-5$\,ks. In other words, just like the
PL$_{\rm norm}$, the brems$_{\rm norm}$  variations are delayed with respect to
the PL spectral slope variations.

Finally, we find that the bremsstrahlung temperature is also variable. A
$\chi^{2}$ test results in  $\chi^{2}=422.1/197$ dof. The kT variations are of
low amplitude with $f_{\rm rms, {\rm kT}}=4.9\pm 0.7$\%, similar to the PL slope
variability amplitude. However, they do not correlate, at any lag, neither with 
brems$_{\rm norm}$ nor with PL$_{\rm norm}$, $\Gamma$ or the observed intensity
variations. 

In summary, our results so far show that: a) Both the PL normalization and slope
are variable, on time scales as short as $0.5-1$ ks, b) The soft excess
component {\it is} variable on short time scales, mainly in amplitude,  c) the
observed flux variations in the soft and hard energy band light curves are
mainly caused by the PL and the soft excess component normalization variations,
d) the PL and bremsstrahlung normalization variations are well correlated, with
no delays, and finally, e) the spectral slope variations seem to lead (by $\sim
1-2$ ks) the PL and bremsstrahlung normalization variations. 

The final step in the investigation of the spectral variations of the source is
to use the best fitting spectral parameter values to determine the fluxes of
the individual components in various energy bands, and then their
cross-correlations. 

\begin{figure*}
\hbox to \hsize{
\psfig{figure=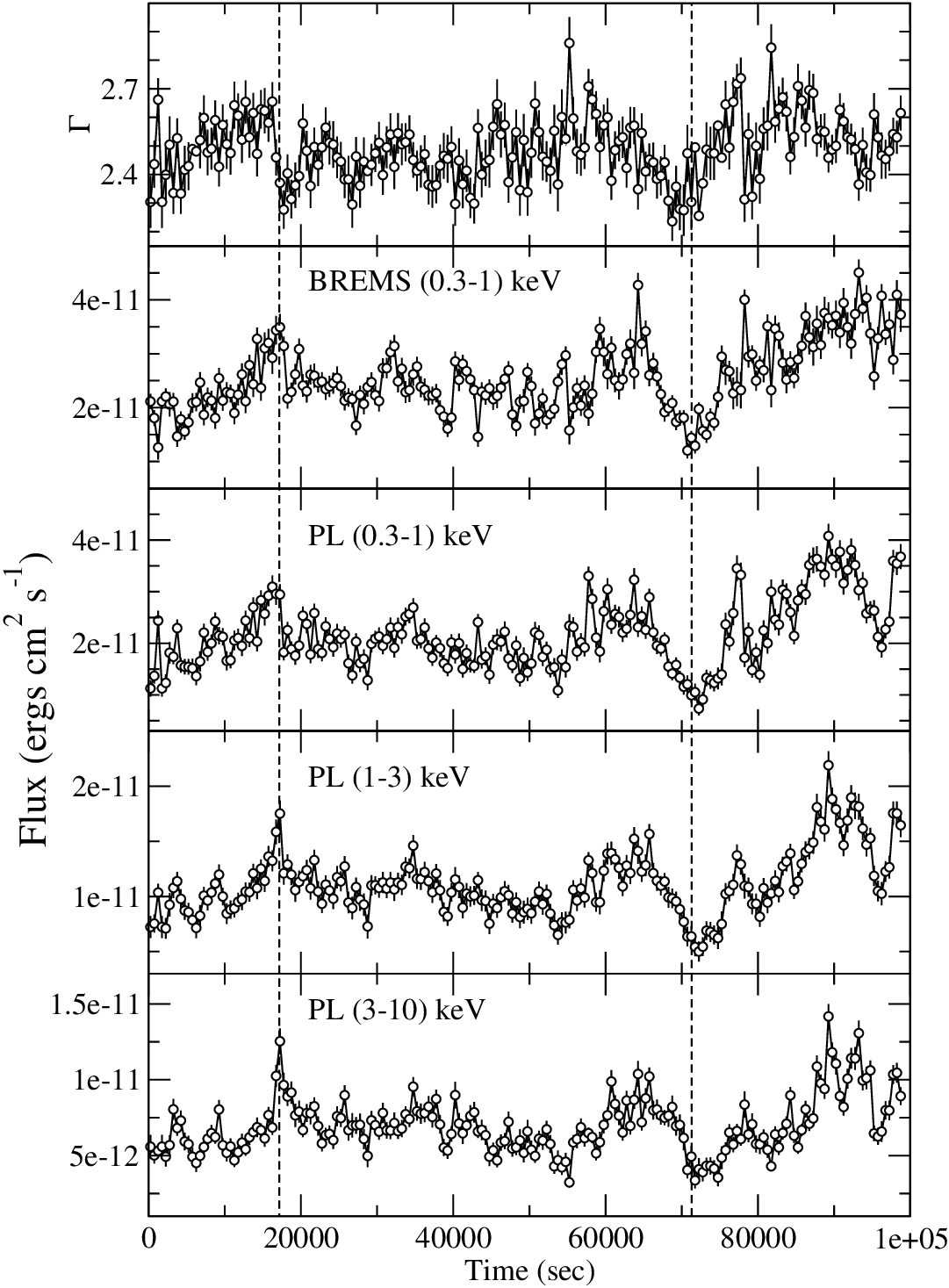,height=10.8truecm,width=8.5truecm,angle=0,%
 bbllx=7pt,bblly=49pt,bburx=528pt,bbury=755pt,clip=}
\psfig{figure=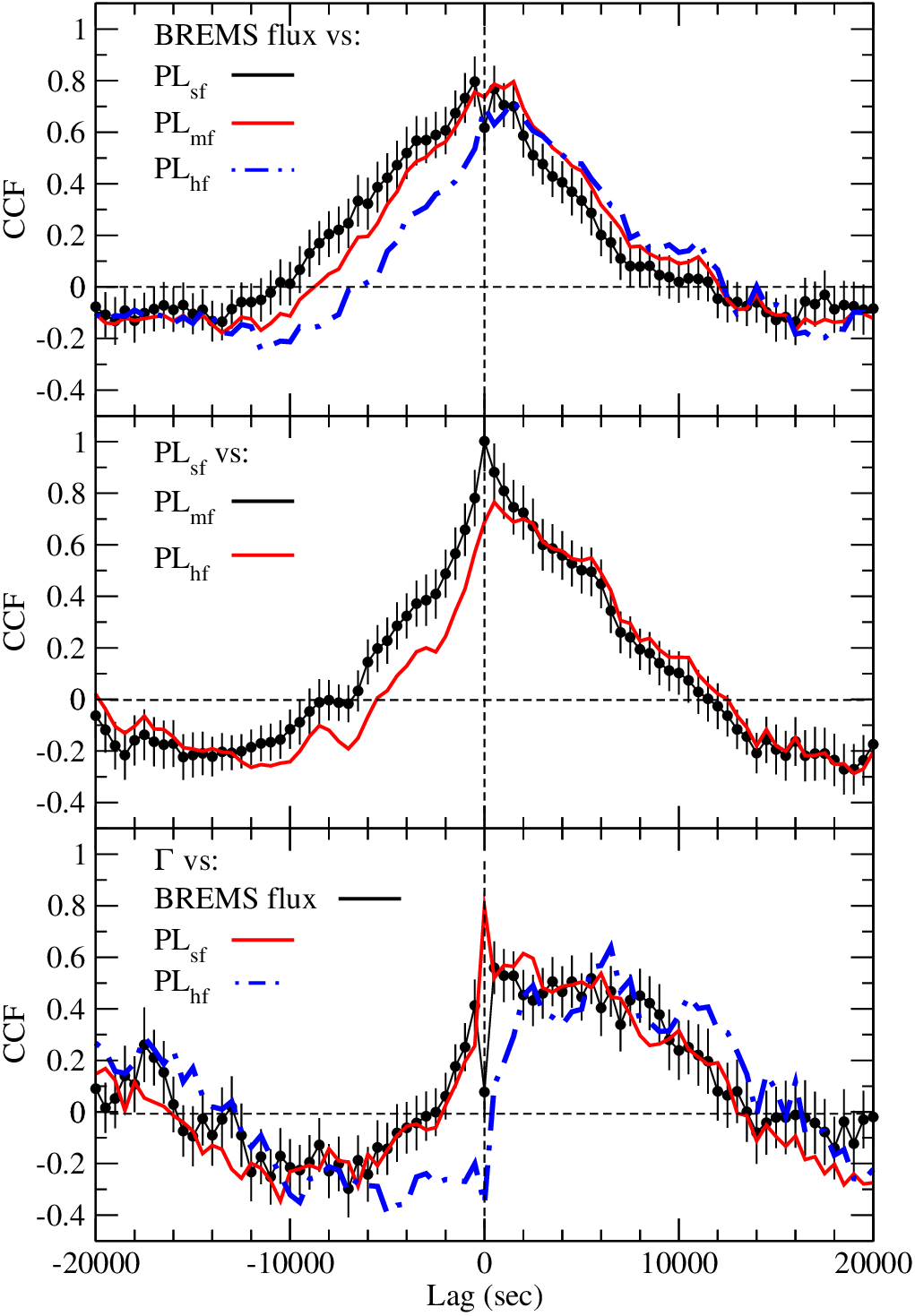,height=10.8truecm,width=8.5truecm,angle=0,%
  bbllx=38pt,bblly=48pt,bburx=529pt,bbury=755pt,clip=} }
 \caption[]{\small Left: The PL best fit spectral slopes (top panel), the
bremsstrahlung $0.3-1$ keV flux (second panel from top), the power law  
$0.3-1$ and $1-3$ keV flux (third and fourth panel from top), and the $3-10$
keV flux (bottom panel), plotted as a function of time. Right:
Cross-correlation functions between the various parameters which are plotted on
the left side.}
\label{fig:09} 
\end{figure*}

\subsection{Flux-flux correlations} 

Fluxes provide a more direct  tracer of the sources' emission characteristics
than the count rates  which are subject to short term spectral variations and
the folding with the detector response. For example, comparing the total
0.3$-$10 keV light curve with the calculated fluxes we find differences around
$\la$ 1\% in a 500\,s bin. This is, however, only a  lower bound for the high
energy band as most of the flux / count rate is found in the soft band.

In the left hand panel of Fig.\,\ref{fig:09}, we plot the bremsstrahlung
$0.3-1$ keV flux, and the PL $0.3-1$, $1-3$ and $3-10$ keV fluxes (PL$_{\rm
sf}$, PL$_{\rm mf}$, and PL$_{\rm hf}$, respectively) as a function of time.
In these plots the estimated errors of the fluxes are generally
smaller than the symbol sizes.
The fractional variability amplitude is $f_{\rm rms,brems}=23.8\pm 0.01$\%, 
$f_{\rm rms,PL_{sf}}=29.5\pm 0.01$\%, $f_{\rm rms,PL_{mf}}=24.5\pm 0.01$\%,
and  $f_{\rm rms,PL_{mf}}=25.1\pm 0.01$\%. The bremsstrahlung flux light curve
shows the smallest amplitude variations, although the differences are not
large.

As before, to ease the visual comparison between them,  we have drawn two
dashed lines to mark the two ``major" events during the first 20 ks of the
observation, and around 70 ks after its start. In this case, it is hard to
judge whether there are any delays between the times that the various band 
flux light curves reach their maximum and minimum values, respectively. 

However, what {\it is} clear is that there are differences in the {\it shape}
of the light curves. While the bremsstrahlung flux increases
steadily for the first $\sim 17$ ks, and then decays sharply within a couple of
thousand seconds, the PL flux light curves show a different behavior. The
difference is enhanced as the energy separation increases. For example, the
PL$_{\rm hf}$ light curve increases sharply within $\sim 2$\,ks before the
bremsstrahlung flux maximum, and then decays smoothly over a period of $\sim 5$
ks. The same behavior is observed in the second ``event''. The bremsstrahlung
flux decays smoothly for over a period of $\sim 5-6$ ks before reaching its
minimum and then rises steadily  immediately after that. The PL$_{\rm hf}$ flux
on the other hand decays much faster (within $\sim 2$ ks), and then increases
at a much slower rate. 

These differences in the rise and decay time scales between the various light
curves are bound to introduce ``delays" between the observed variations.
Indeed, on the top right hand panel of Fig.\,\ref{fig:09}, we show the CCF
between the  bremsstrahlung and the  PL$_{\rm sf}$, PL$_{\rm mf}$, and
PL$_{\rm hf}$ light curves. The bremsstrahlung flux vs PL$_{\rm sf}$ CCF shows
a maximum at almost zero lag.  However, the bremsstrahlung vs PL$_{\rm
hf}$ and vs PL$_{\rm hf}$ CCFs are heavily skewed towards positive lags, with
broad maxima at lags $\sim 0-2$ ks. These ``delays" are caused to a large
extend  by the differences in the rise and decay time scales of the ``flares"
in the respective light curves. 

In the middle right hand panel of Fig.\,\ref{fig:09} we plot the PL$_{\rm sf}$
vs PL$_{\rm mf}$ and the PL$_{\rm sf}$ vs PL$_{\rm hf}$ cross-correlation 
functions. In agreement with the CCFs plotted in the panel above, they also 
look asymmetric, and shifted to positive lags. As before, these ``delays" are
mainly caused by the differences in the shape of these light curves. 

\begin{figure}
\psfig{figure=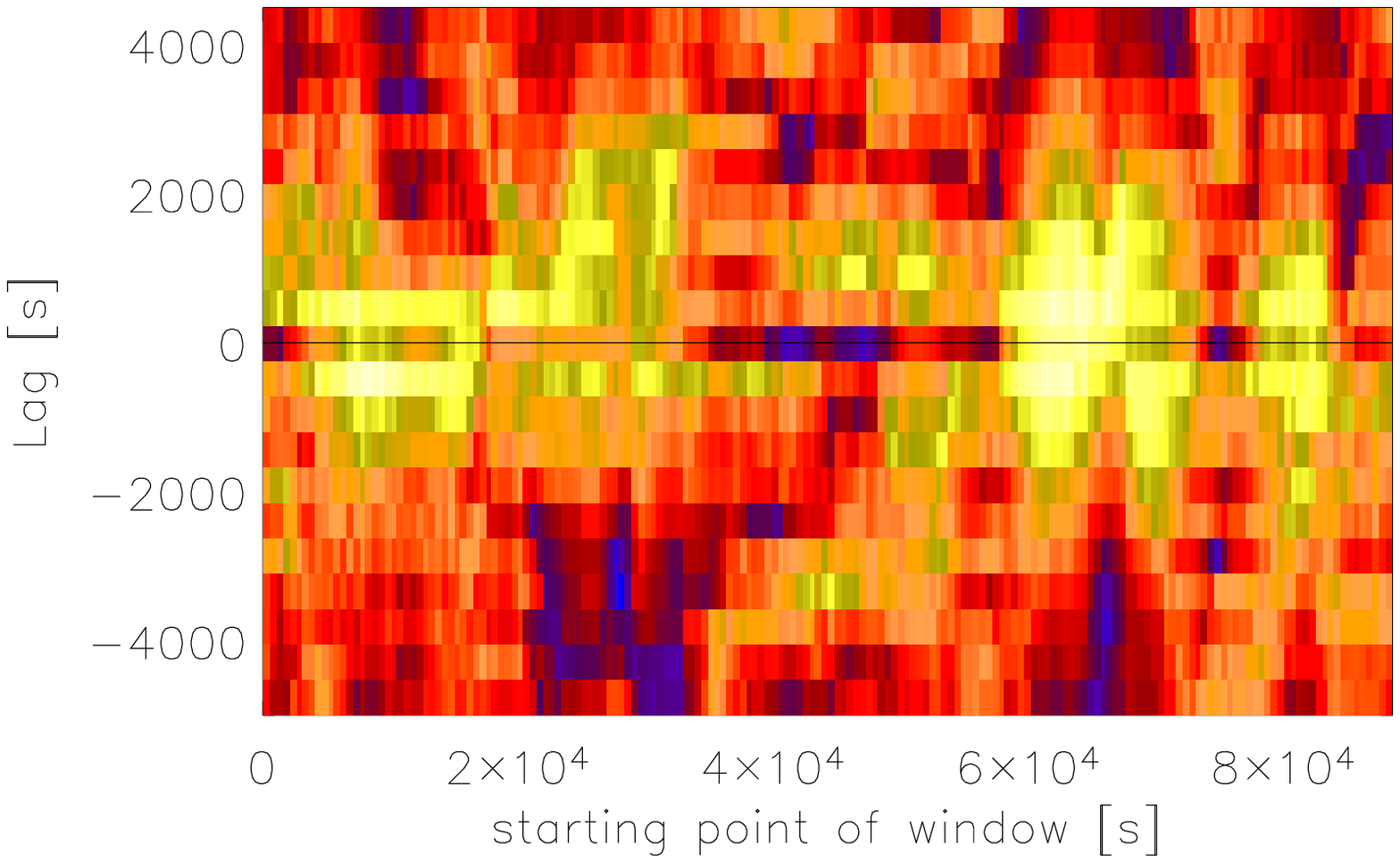,height=4.1truecm,width=8.5truecm,angle=0,%
 bbllx=2pt,bblly=48pt,bburx=490pt,bbury=290pt,clip=}
\psfig{figure=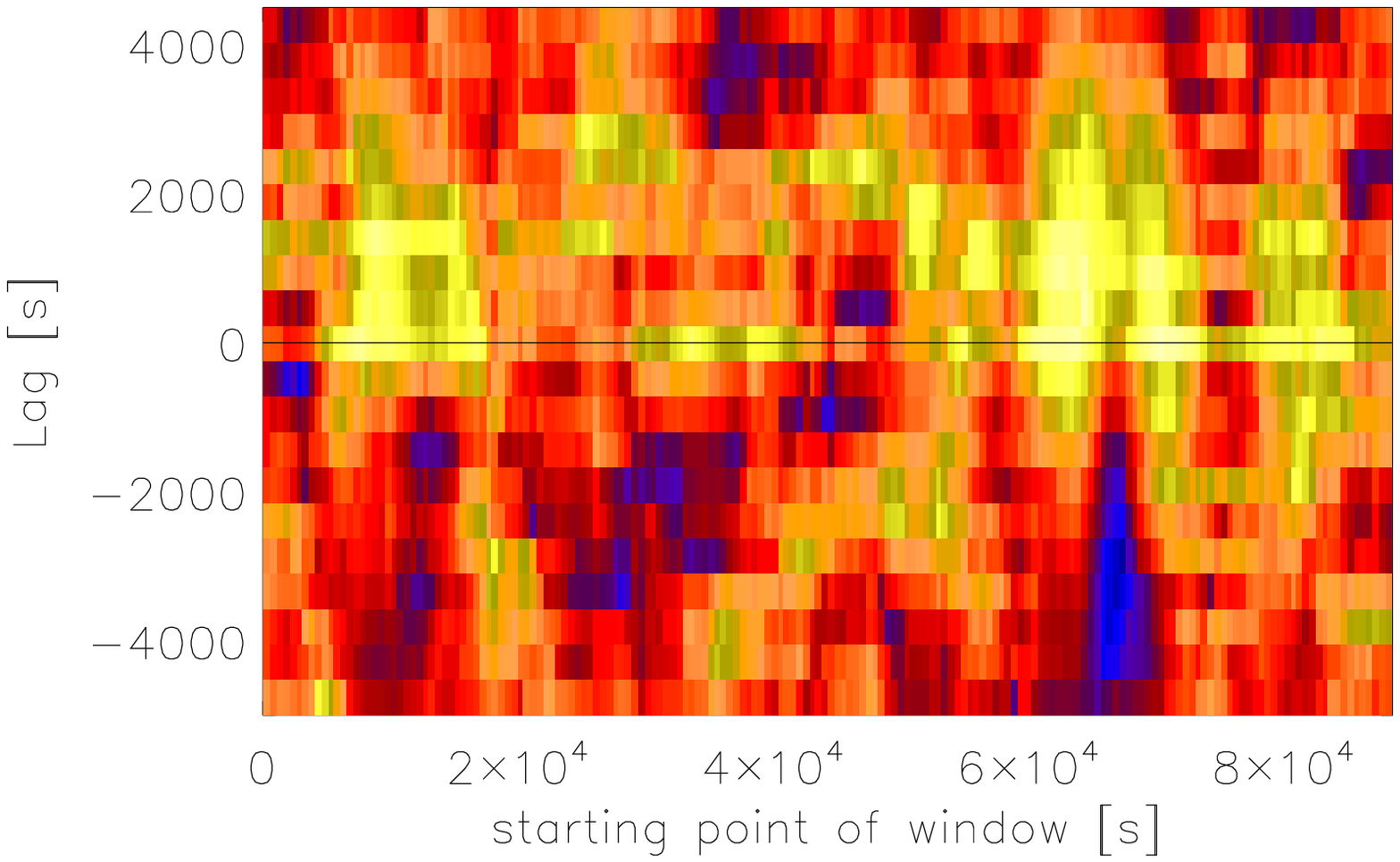,height=4.1truecm,width=8.5truecm,angle=0,%
    bbllx=2pt,bblly=48pt,bburx=490pt,bbury=287pt,clip=}
\psfig{figure=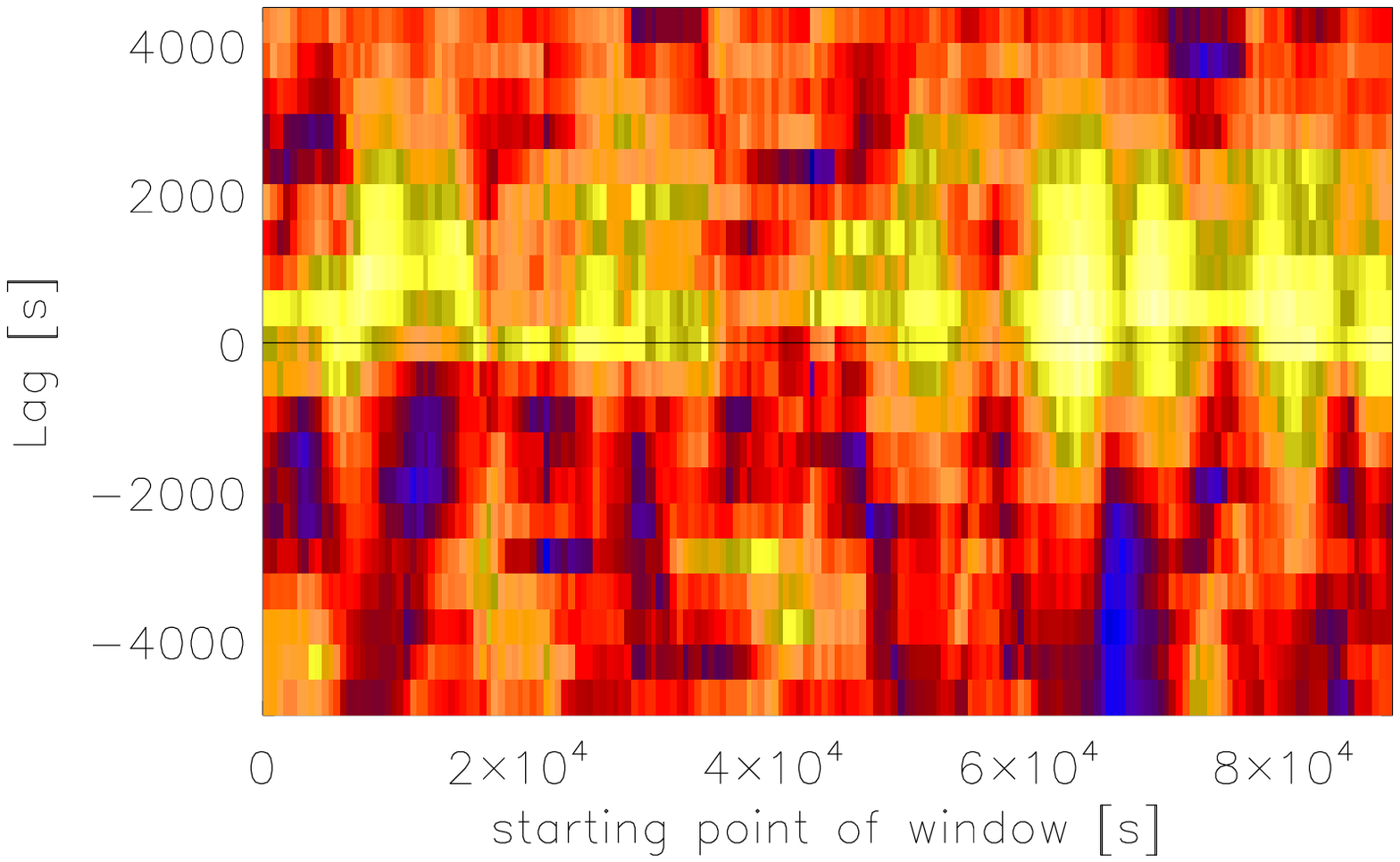,height=4.7truecm,width=8.5truecm,angle=0,%
          bbllx=2pt,bblly=1pt,bburx=490pt,bbury=287pt,clip=}
\caption[]{\small Sliding window CCFs between the bremsstrahlung and soft power law flux
    (upper panel), between bremsstrahlung and hard power law flux (middle panel) and 
    between soft  and hard band power law fluxes (lower panel). The time binning
    is 500\,s, the window length 
    $\mathcal{L}$ = 10\,ks; color coding and description as in Fig.\,\ref{fig:slide1}.} 
\label{fig:slide2}
\end{figure}

The lower than unity CCF maximum values are caused by the fact that the
correlation coefficient changes on time scales of a few ks. This is
demonstrated in Fig.\,\ref{fig:slide2} which shows on top the sliding window
CCF of the bremsstrahlung and the soft power law flux.  The correlation
coefficient changes with time and shows its maxima at the two major flux
variations events we have discussed before. In between, most probably the low
amplitude variations do not allow us to pick any significant signals. In the
middle and bottom panels we show  the sliding window CCF of the bremsstrahlung
and the medium power law flux and the soft vs hard power law flux,
respectively. In both cases, the variations in the  softer energy band lead
those in the higher energy band. The maximum CCF signals appear at the two
prominent flux variability events, while no significant signals appear in
between.

\subsection{Spectral slope vs Flux correlations} 

The top left hand panel of Fig.\,\ref{fig:09} displays  the   power law  slope
$\Gamma$ as a function of time. The curve looks roughly similar to the flux
light curves, shown below, although it is not as variable. Looking again at the
two main ``events" that we have mentioned above, it now becomes clear that the
$\Gamma$ variations are {\it leading} the flux variations of both the
bremsstrahlung and PL components, in all energy bands. 

In the bottom right hand panel of Fig.\,\ref{fig:09}, we show the CCF between
$\Gamma$ and the bremsstrahlung,  the PL$_{\rm sf}$, and  the 
PL$_{\rm hf}$ flux light curves.  They all show broad maxima shifted towards
positive lags. The degree of the asymmetry increases with increasing energy,
and is maximal in the  $\Gamma$ vs PL$_{\rm hf}$ CCF. This result demonstrates
clearly that in \ark the PL spectral slope variations are leading the flux
variations. 

\begin{figure}
\psfig{figure=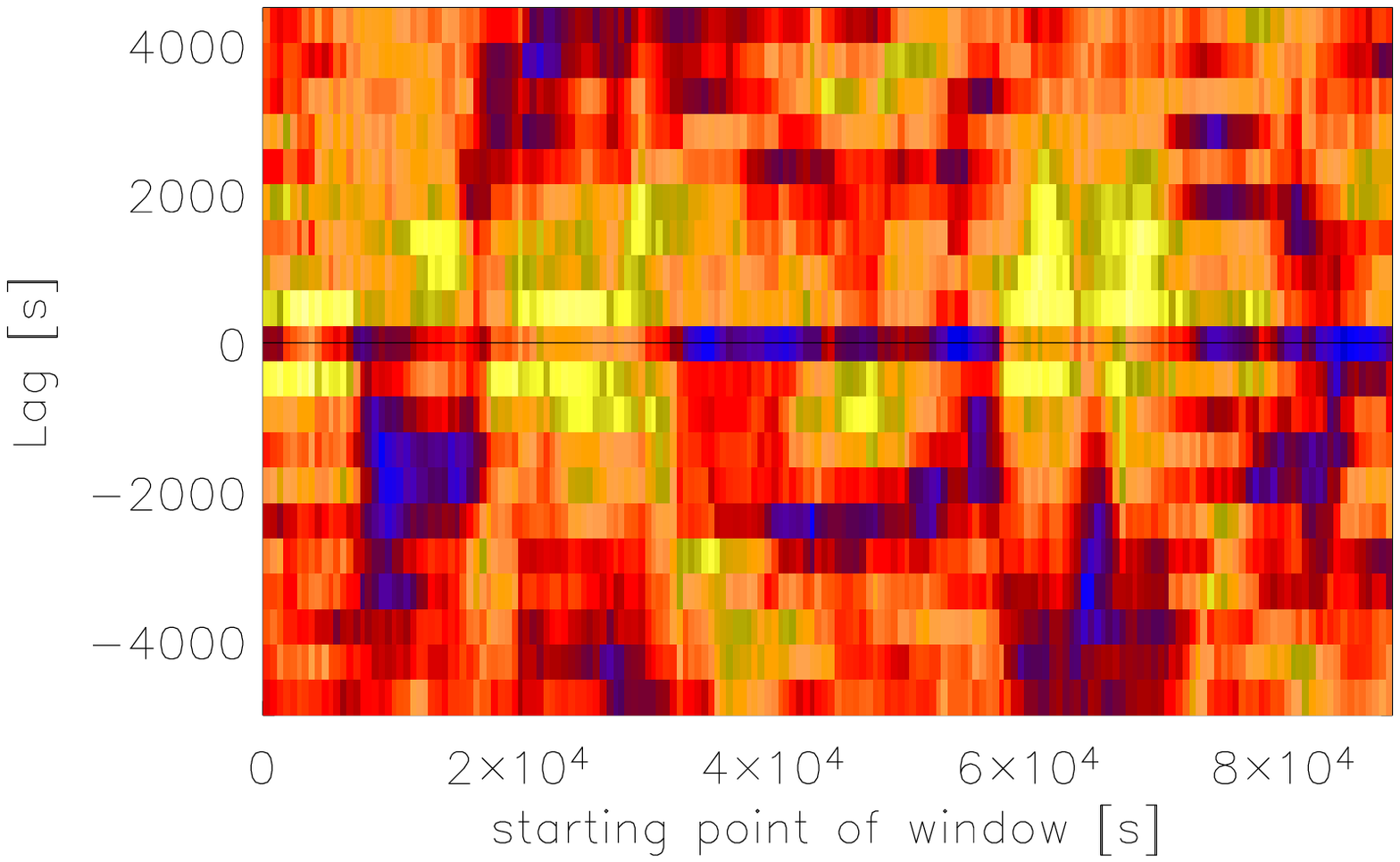,height=4.1truecm,width=8.5truecm,angle=0,%
 bbllx=2pt,bblly=48pt,bburx=490pt,bbury=290pt,clip=} 
\psfig{figure=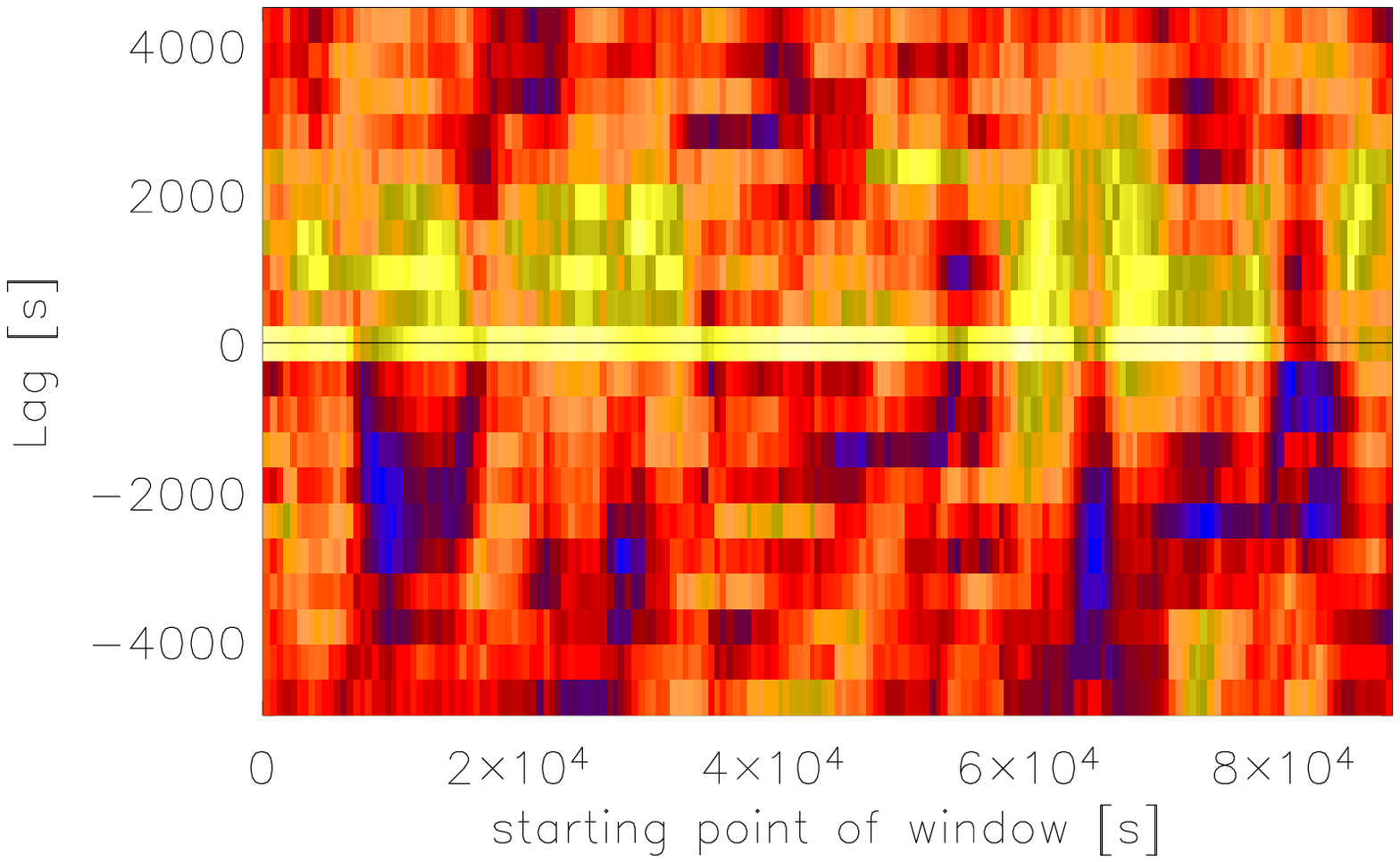,height=4.7truecm,width=8.5truecm,angle=0,%
                bbllx=2pt,bblly=1pt,bburx=490pt,bbury=287pt,clip=}                  
\caption[]{\small Sliding window CCFs between the photon index $\Gamma$ and the 
   bremsstrahlung flux (upper panel)  and  between $\Gamma$ and the soft power law flux
   (lower panel). Color coding and description as in Fig.\,\ref{fig:slide1}.}
\label{fig:slide3}                                                                                                            
\end{figure}

The CCF maximum values are rather low, being  roughly equal to $0.6$. This,
together  with broadness of the CCFs and their ``noisy'' structure can be
explained by the fact that, on closer inspection, the $\Gamma$ and flux
variations are not always in phase (see also Fig.\,\ref{fig:corr}), especially
towards the end of the observation where $\Gamma$ remains roughly constant and
does not follow the flux variations. Furthermore, as before, the sliding
window CCFs (Fig.\,\ref{fig:slide3}) shows that changes in the power law index
$\Gamma$ and the bremsstrahlung/power-law flux are intimately related during
the two major variability events, but in other times the signal simply fades
away.

\section{Discussion}
\smallskip

In this paper we present the results from a detailed spectral variability study
of \ark using the data from a recent, 100 ks long, \xmm observation of the
source. We have estimated the cross-correlation functions between the soft and
hard energy band light curves, and between these light curves and  conventional
hardness ratios. Furthermore, due to the high signal-to-noise ratio of the
data, we were able to extract high quality,  individual spectra from data
stretches of 500 s each. We were able to study, for the first time in the case
of \ark, the spectral variability properties of the source down to a time scale
as short as 500\,s, by fitting  a power law plus bremsstrahlung model to these
spectra. 

Ark\,564 is an ideal AGN for
a spectral variability investigation. First of all it is highly variable (i.e.
it displays large amplitude variations on short time scales), and bright, hence
we are able to probe accurately its high frequency variations.  Furthermore, as
we have shown in Paper I,  its X-ray spectrum is almost ``clean" of any
significant absorption and/or emission features.  For example, the top panel of
Fig.\,12 in Paper\,I demonstrates that the fraction  of any manifestations of a
warm absorber is no more than $4$\% percent  of the X-ray continuum flux  in
the energy band between $\sim 0.7-0.9$ keV.  Furthermore, the comparison of the
time-averaged spectrum of the source with that of 3C 273 (shown in Fig.~8 of
Paper I) shows clearly that the overall spectrum of Ark 564 is smooth and is
mainly characterized by broad continuum spectral components, which are
unaffected by strong  absorption and/or emission features. As a result, it is
possible to determine the X-ray continuum with the use of simple
phenomenological models, without any ambiguity arising from the presence of
significant distortions due to strongly absorbing and/or emitting components 
and subsequently study its variations even on time scales  as short as a few
hundred seconds.

We use a simple PL model to account for the emission above 3 keV of the source.
This fits well all the spectra, but only as long as $\Gamma$ varies by up to
$\sim 0.4$.
There have been cases in the past few years  where the observed
spectral variations of a few sources have been interpreted in terms of a
``two-component" model (i.e. Ponti \etal 2006, and references therein). 
According to this model,  the observed flux variations are caused by a
constant-slope power law which varies in normalization only. The spectral shape
variations are artificial and are  introduced by the interplay of this and a
second spectral component  (usually attributed to ionized reflection by the
disc) which is almost constant (in flux and shape). To test this hypothesis in
the case of \ark, we fitted the lowest and highest flux source spectra with a
PL plus the ionized reflection model REFLION of Ross \& Fabian (2005). We kept
the values of the reflection component parameters fixed to those found in Paper
I (as this component is supposed to be constant). The source spectra above 1
keV can be fitted well by this model (at lower energies, we need and extra
spectral component to fit the high flux spectrum, just like in the case of the
time-averaged spectrum). However, this is possible only if the PL slope changes
by a factor of 0.6$\pm 0.13$. This is comparable (in fact, even slightly
larger) to the $\Delta \Gamma$ we detect when we parameterize the spectra with
the simple  PL+bremsstrahlung model. We believe that the detected spectral
slope variations in the hard band of \ark are an intrinsic property of the
source. 

Regarding the soft band spectrum, we have used a simple bremsstrahlung law to
account for the soft excess emission of the source. This does not imply that we
believe the soft-excess  component is indeed bremsstrahlung emission from a hot
plasma (see below). However, to the extend that a variable PL component fits
well the 3-10 keV band and extends to slower energies as well (as is shown by
the study of the time-averaged spectrum in Paper I) then  the facts that:  a)
the soft X-ray continuum is smooth and almost featureless and b) the
bremsstrahlung model fits it well, guarantee that the flux estimates (at least)
of the soft component  must be similar to what we find, irrespective of what
models one may use to fit the data. 

In Paper\,I the possible detection of an absorption line at 
$\sim 8.1$ keV in the time average spectrum of the source  was
discussed. If real, this would suggest the presence of highly ionized       
out-flowing gas with a column density of $\sim 10^{23}$ cm$^{-2}$.
Significant variations in the column density, covering factor and/or        
ionization state of the gas could result in spectral variations, even at    
energies above $2-3$ keV.
 
We fitted some of the 500 s long spectra with a  ``power law  
plus Brems plus \footnotesize{ABSORI}" model but it was not possible to     
constrain both the N$_{\rm H}$ and ionization parameter ($\xi$) of the      
gas. Even when N$_{\rm H}$ was kept frozen to a constant value, the         
uncertainty on $\xi$ was very large.

To investigate then the possible effects of $\xi$ variations we used        
the \footnotesize{XSPEC} command \footnotesize{FAKE} and created
synthetic PN spectra (500 s long each) assuming a PL plus  a  
bremsstrahlung and an \footnotesize{ABSORI} model component with N$_{\rm    
H}=10^{23}$ cm$^{-2}$. The PL and Brems normalizations as well as the       
$\xi$ values were varied and the resulting spectra were fitted with a     
power law  for energies above 2 keV. Due to the limited signal 
to noise of the    
500 s spectra, they could all be well fitted by the PL model.

It seems that the gas must be higher ionized than what the model's 
upper limit of $\xi=5000$ implies (due to the presence of model spectral    
features in the soft energy band which we do not observe in reality),       
but it is not certain whether  $\xi$ variations will then result in         
significant $\Gamma$ variations. Nevertheless it seems not impossible       
that $\xi$ variations can cause spectral variations similar to what we      
observe in \ark. We did not consider the cases of N$_{\rm H}$ or
covering fraction variations, as the probability that these parameters      
vary on time scales as short as less than an hour is rather small.

However, our results regarding the flux variations in various energy
bands and their correlations (see Sec.~4.3) should not depend on the        
model chosen to fit the 500 s long spectra, as long as it fits them         
well. It further seems rather difficult to explain  why the $3-10$ keV
band flux   variations should be delayed with                             
respect to the $1-3$ keV band variations if most of the spectral
variations are caused by changes in the properties of the warm absorbing    
material, which should affect the whole spectrum simultaneously. It is     
also difficult to understand in this case why variations of the 
slope $\Gamma$ would lead the flux variations in all bands.
  
In summary, we believe that most of our results, which we summarize below,
do not depend on the choice of the models we have used to parameterize the
source's spectral variations and that they are representative of the intrinsic
properties of the source's emission components. In particular, the results
regarding the flux variations in various energy bands and their correlation
with the observed spectral shape variations  do not depend on the way we model
the source's spectrum.

\subsection{Summary of the results}
 
Our findings can be summarized as follows:
 
1) The soft ($0.3-1$ keV) and hard ($3-10$ keV) band light curves are well
correlated at zero lag. The CCF is asymmetric, indicating that the hard band
photons are delayed, in some way, with respect to the soft band photons. 

2) The hard/soft band ratio is variable, but its variations are not well
correlated with those observed in the individual light curves.  

The first result is consistent with the results of Arevalo \etal\ (2006), who
use cross-spectral techniques to study in detail the correlation between the
soft and hard band light curves. However, the interpretation is difficult
because {\it both} the soft and hard band  spectral components may contribute to
the soft band light curve. Our time  resolved spectral analysis 
can resolve this issue.  The results from the application of this
method to the  \ark\ data are as follows: 

3) The shape of the hard band spectrum changes significantly on time scales as
short as $\la 1$\,ks. These changes can be well described by power law slope
changes of the order of $\Delta \Gamma \la 0.4$, maximum.

4) An excess emission above the extrapolation of the PL model to energies down
to $0.3$ keV is always present. A bremsstrahlung model fits this component well
at all times. 

5) The soft excess flux is significantly variable, even on the shortest time
scales of $\sim 0.5-1$\,ks that we can probe. 

6) The PL and bremsstrahlung normalization variations are the prime drivers of
the observed intensity variations of the source.  The variations of the two
components are well correlated with each other. 

7) The bremsstrahlung  $0.3-1$ keV, and the PL $0.3-1$, $1-3$, and  $3-10$ keV
flux light curves have different ``shapes". The rise and decay time scales
decrease with increasing energy. Consequently, the PL, $3-10$ keV flux light
curve is delayed with respect  to {\it both} the bremsstrahlung and the PL,
$0.3-1$ keV flux light curves. This result explains the CCF results between the
$0.3-1$ keV and $3-10$ keV band light curves. And, finally, 

8) The $\Gamma$ variations are always preceding those in the bremsstrahlung and
PL component flux light curves. The delays increase with increasing energy.
Such a result could not be detected in the CCF between the hardness ratio and
the intensity light curves, because, as it is defined, the hardness ratio is not
representative of the slope of the PL component. 

We discuss below some possible implications of these results in the context of
a few current theoretical models.

\subsection{Comptonization models}

It is widely accepted that the high energy, X-ray power law continuum in AGN 
is  produced  in a hot corona which is located above the accretion disc by
Comptonization of soft thermal photons (e.g. Haardt \& Maraschi 1993, Haardt
\etal 1994). This model is in particular attractive for NLS1 galaxies as
reflection features  (like the iron line) tend to be more suppressed by Compton
scattering in  the corona itself (Matt \etal 1997), and a magnetically heated
corona, where  the energy is primarily stored in magnetic fields which
reconnect  and release their energy in flares (Merloni \&  Fabian 2001), can
account for the strong variability of these sources (Brinkmann \etal 2004).

The spectral study of the time-averaged spectrum of the source (Paper I) has
shown that, if thermal Comptonization is involved in the production of X-rays in
\ark, the expected high energy cut-off is located at energies far above the
energy range of the XMM instruments. In this case we cannot constrain the models
as, in the XMM energy band, the measured power law slope is merely an indicator 
for a wide range of possible combinations of the temperature kT and the optical
depth $\tau$ of the scattering corona (Titarchuk \& Lyubarskij 1995). 

However, the results from the present work can place constraints on simple
Comptonization models. For example, the $\Delta \Gamma$ variations that we
observed can be compared with the predictions of a simple Comptonization model,
where X-rays are produced in a single corona with a uniform temperature and
optical depth. Let us consider, for instance, the strong flux decrease from 
$\ma$ 65\,ks in the light curve and the  corresponding changes of the power law
component (Fig.\,\ref{fig:gamfit}): while the power law slope gets flatter
($\Gamma$ changes from $\sim$ 2.6 to $\sim$ 2.3) the flux (normalization)
decreases  by nearly a factor of 2. For a fixed optical depth, an increase of
the corona temperature  will lead to a flatter  power law slope but to an
increase of the emitted flux as well, in contrast to the  observations.
Similarly, increasing the optical depth by the amount required to account for
the slope changes at constant temperature will, again, lead to an increase of
the flux. Obviously, a simple one-parameter change of the physical  conditions
of the emission region cannot describe the observations. Either several
parameters change simultaneously in a  complex way -  which would be natural for
a physically evolving system (and models of this type have been discussed by
e.g. Haardt \etal 1997), or the emission conditions  change on very short time
scales (note that the current Comptonization models are usually ``equilibrium''
models). 

Focusing on the flux increase during the first $15-17$ ks after
the start of the observation one could think of a coronal area storing
lots of energy in magnetic fields, which is subsequently released and suddenly
heats up electrons. There are initially few
photons around to cool the electrons, so  there is a hot, under luminous
corona giving a flat but weak power law continuum. After some cooling time
(which is of the order of the crossing time through the corona), reprocessing 
in the disc allows more seed photons which results in a corona with lower
temperature (Haardt, private communication).
The power law slope increases (as observed), together with its
flux, as the cooling of the corona is more efficient. 
In such a case one would expect the spectral slope to lead the PL and 
bremsstrahlung normalization
increase, as observed. However, at the same time, as the temperature decreases,
we would also expect the PL hard band to  increase `before', or `less sharply',
than the  soft band flux, opposite to what is observed. Presumably, one has to
take into account both the (unknown mechanism of the) coronal heating and 
cooling, but this task is beyond the aims of the present work.

The various time scales we found in the analysis provide perhaps 
important clues for the geometry of the X-ray sources in AGN. We find that  flux
and spectral variations propagate from the  soft to the hard energy bands with
typical delays of $\sim 1-2$\,ks. The light crossing time scale of the emission
region is, however, only of the order of $\la$ 200\,s. Therefore, these delays
are longer than the time scale for Compton - up-scattering of the photons
in a corona of moderate optical depth. One could think of multiple emission
regions, perhaps with different temperatures and/or optical depths. In this case,
propagation effects of any disturbances, that first affect the soft and then the
hard band emitting regions cannot be ignored. For example, models in which
inwardly-propagating variations in the local mass accretion rate affect the
X-ray producing region, and relatively harder X-ray bands are associated with
emissivity profiles that are more centrally concentrated (e.g. Arevalo \& Uttley
2006, and references therein) could explain, qualitatively, the propagation of
the flux variations from the soft to the hard bands (i.e. the CCFs between the
PL flux light curves in the various energy bands) that we observe. 

\subsection{The soft excess component}

Although the soft excess component in the individual 500 s long spectra is
always well fitted by a bremsstrahlung model,  we believe that this  is probably
only an 'acceptable' representation of a more complex   
 emission from a hot, thermal plasma. The simplest
reason for this is that a bremsstrahlung model alone does not provide an 
acceptable fit to the
time-averaged spectrum as well (Paper I). 

In this work we show conclusively that the soft excess component in \ark is
variable on time scales as short as $\sim 0.5-1$ ks. For a $2-3\times 10^6
M_\odot$ black hole mass, the dynamical time scale at a distance of 3
Schwarzschild radii is $\sim 200-300$ sec. Consequently, it is conceivable that,
from the variability's point of view, the soft excess component can represent
direct thermal emission from the innermost region of the accretion disc. The
first argument against this possibility is the fact that a black-body emission
did not provide a fit better than the simple bremsstrahlung model, as we would
expect in this case. 

Czerny et al. (2003), and Gierlinski \& Done (2004) have shown that, if the soft
excess in AGN can be modeled by black body emission, then the resulting
temperature is the same in many AGN, irrespective of the mass of their black
hole. This is not what one should expect to observe.   In agreement with these
results, it was argued in Paper I that the soft excess component in \ark cannot
represent thermal emission from the disc because its temperature is too high for
a standard $\alpha$ - disc. 

This is a model dependent argument, but the results from the present work
provide some more arguments against the possibility that 
the soft X-ray component in \ark represents  thermal emission
from the disc. For example, we find that the soft X-ray flux is highly 
correlated with the PL
flux (see Section 4.3). In fact, the soft excess variations lead those in the PL
flux light curves. The obvious explanation is that the soft excess photons are
the seed photons for the hard X-ray power-law. However, in this case we would
expect the PL flux light curves to be ``smoother" and of lower amplitude than
the soft excess light curve. 
As this is opposite to what we observe we do
not believe that it is a plausible scenario.

A second possibility for the soft excess component is that it represents ionized
smeared reflection from the accretion disc (Crummy et al. 2006). A signature of
ionized reflection is the excess emission which occurs in the 0.2-2 keV band
owing to lines and bremsstrahlung from the hot surface layers.  The study of the
time-averaged spectrum of \ark in Paper I showed that this is can be plausible
scenario for the source. If true, then our results show that the soft excess
emission in this source responds very fast to the primary component variations.
This is opposite to what has been observed in other AGN, based on the fact that
the iron line usually does not respond to the continuum variations. 

According to this scenario, since the heating of the disc  surface layers is
mainly caused by the absorption of relatively low energy photons (i.e. $\la 3$
keV) of the X-ray continuum, it is reasonable to expect that the soft component
and the $0.3-1$ keV PL light curve will be well correlated, as is observed. In
fact, one may even notice a slight asymmetry towards negative lags in the soft
excess vs the  $0.3-1$ keV, PL flux CCF (filled circles in the upper, right hand
panel in Fig.~9) which is expected due to the light-travel time delay between
the detection of the  X-ray source and the disc's response signals. In this
case, the correlation of the soft excess flux with the the PL flux at higher
energies must be mainly a secondary effect, caused by the intrinsic correlation
of the soft with the higher energy PL bands. 

Furthermore, using the REFLION model in XSPEC, we produced two, 500 s long 
synthetic spectra with a PL component of $\Gamma=2.3$ and $\Gamma=2.7$,
respectively, and a reflection component with solar abundances. We assumed that
the overall PL flux was 4 times higher in the second case,  and as a result, the
reflected component flux and the ionization parameter has also increased by the
same factor. The two spectra had an average count rate similar to the \ark count
rate in the low and high flux state, respectively. We then fitted the resulting
spectra with a simple bremsstrahlung plus a PL model, and saw that the best
fitting kT values were very similar in both cases. This is in agreement with our
results that kT shows small amplitude variations, and does not correlate with
any other model parameter. Perhaps, the $\Gamma$ and flux variations in Ark
564 vary in such a way so that the shape of the soft excess component stays
roughly constant during the intensity variations. 
 
Concluding, one can say that the excellent data quality, the relatively
``clear'' emission pattern of \ark and the variety of the above 
mentioned emission scenarios certainly asks for a more detailed
study of the theoretical models for the high energy emission from NLS1
galaxies.

\vskip 0.4cm
\begin{acknowledgements}
This work is based on observations with \xmm\, an ESA science mission
with instruments and contributions directly funded by ESA Member States
and the USA (NASA). The \xmm project is supported by the        
Bundesministerium f\"ur Wirtschaft und Technologie/Deutsches Zentrum
f\"ur Luft- und Raumfahrt (BMWI/DLR, FKZ 50 OX 0001), the Max-Planck
Society and the Heidenhain-Stiftung.We gratefully acknowledge travel support through
the bilateral Greek-German IKYDA project based personnel exchange program.    
\end{acknowledgements}

\end{document}